\documentclass{article}
\usepackage{spconf,amsmath,epsfig,listings,rotating}

\newcommand{\Octave}{\textit{Octave}}
\newcommand{\Matlab}{\textit{Matlab}}
\newcommand{\OctFile}{\textit{oct-file}}
\newcommand{\OctFiles}{\textit{oct-files}}

\title{Sparse Matrix Implementation in Octave}
\name{David Bateman$^\dag$, Andy Adler$^\ddag$}
\address{$\dag$ Centre de Recherche, Motorola\\
	Les Algorithmes, Commune de St Aubin\\
	91193 Gif-Sur-Yvette, FRANCE\\
	email: David.Bateman@motorola.com \\
	\\
	$\ddag$ School of Information Technology and Engineering (SITE)\\
	University of Ottawa\\
	161 Louis Pasteur\\
	Ottawa, Ontario, Canada, K1N 6N5\\
	email: adler@site.uottawa.ca}

\begin{document}

\maketitle

\begin{abstract}
This paper discusses the implementation of the sparse matrix support
with {\Octave}. It address the algorithms that have been used, their
implementation, including examples of using sparse matrices in
scripts and in dynamically linked code. The octave sparse functions
the compared with their equivalent functions with {\Matlab}, and
benchmark timings are calculated.
\end{abstract}

\section{Introduction}
\label{sec:intro}

The size of mathematical problems that can be treated at any particular
time is generally limited by the available computing resources. Both
the speed of the computer and its available memory place limitations on
the problem size.

There are many classes of mathematical problems which give rise to
matrices, where a large number of the elements are zero. In this case
it makes sense to have a special matrix type to handle this class of
problems where only the non-zero elements of the matrix are
stored. Not only does this reduce the amount of memory to store the
matrix, but it also means that operations on this type of matrix can
take advantage of the a-priori knowledge of the positions of the
non-zero elements to accelerate their calculations. A matrix type that
stores only the non-zero elements is generally called sparse.  

This article address the implementation of sparse matrices within
{\Octave}~\cite{octave.website,eaton.2003}, including their storage, creation,
fundamental algorithms used, their implementations and the basic
operations and functions implemented for sparse matrices. Benchmarking
of {\Octave}'s implementation of sparse operations compared to their
equivalent in {\Matlab}~\cite{matlab.gettingstarted} are given and their 
implications discussed. Furthermore, the method of using sparse
matrices with {\Octave} {\OctFiles} is discussed.

\begin{figure}
  \begin{minipage}[b]{1.0\linewidth}
    \centering
    \centerline{\epsfig{figure=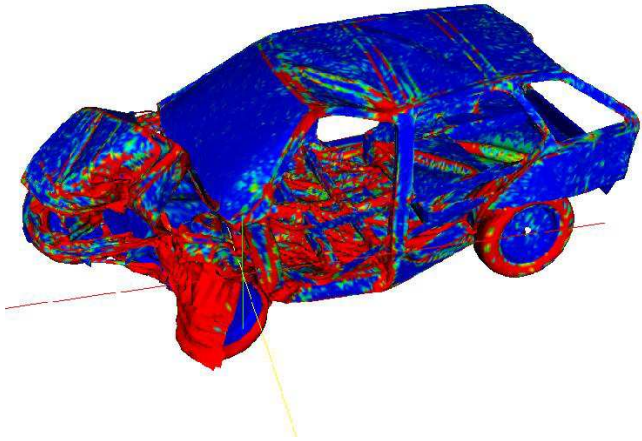,width=8.5cm}}
  \end{minipage}
  \begin{center}
  \caption{Image of automobile crash simulation, blue regions indicate
  rigid-body behaviour. Image courtesy of BMW and Fraunhofer Institute
  SCAI.}
  \label{fig:bmw_image}
  \end{center}
\end{figure}

In order to motivate this use of sparse
matrices, consider the image of an automobile
crash simulation as shown in Figure~\ref{fig:bmw_image}.
This image is generated based on ideas of DIFFCrash
\cite{diffcrash} -- a software package for the
stability analysis of crash simulations.
Physical bifurcations in automobile design and numerical
instabilities in simulation packages often cause extremely
sensitive dependencies of simulation results on even the
smallest model changes.
Here, a prototypic extension of DIFFCrash uses octave's
sparse matrix functions (and large computers with lots
of memory) to produce these results.

\section{Basics}
\label{sec:basics}

\subsection{Storage of Sparse Matrices}
\label{sec:storage}

It is not strictly speaking necessary for the user to understand how
sparse matrices are stored. However, such an understanding will help
to get an understanding of the size of sparse matrices. Understanding
the storage technique is also necessary for those users wishing to
create their own {\OctFiles}.

There are many different means of storing sparse matrix data. What all
of the methods have in common is that they attempt to reduce the
complexity and storage given a-priori knowledge of the particular
class of problems that will be solved. A good summary of the available
techniques for storing sparse matrix is given by
Saad~\cite{saad.sparskit}.  With full matrices, knowledge of the point
of an element of the matrix within the matrix is implied by its
position in the computers memory. However, this is not the case for
sparse matrices, and so the positions of the non-zero elements of the
matrix must equally be stored.

An obvious way to do this is by storing the elements of the matrix as
triplets, with two elements being their position in the array 
(rows and column) and the third being the data itself. This is conceptually
easy to grasp, but requires more storage than is strictly needed.

The storage technique used within Octave is the compressed column
format.  In this format the position of each element in a row and the
data are stored as previously. However, if we assume that all elements
in the same column are stored adjacent in the computers memory, then
we only need to store information on the number of non-zero elements
in each column, rather than their positions. Thus assuming that the
matrix has more non-zero elements than there are columns in the
matrix, we win in terms of the amount of memory used.

In fact, the column index contains one more element than the number of
columns, with the first element always being zero. The advantage of
this is a simplication in the code, in that their is no special case
for the first or last columns. A short example, demonstrating this in
C is.

{\small
\begin{lstlisting}
  for (j = 0; j < nc; j++)
    for (i = cidx (j); i < cidx(j+1); i++)
       printf ("Element (%i,%i) is %d\n", 
          ridx(i), j, data(i));
\end{lstlisting}
}

A clear understanding might be had by considering an example of how the
above applies to an example matrix. Consider the matrix

\begin{center}
  \begin{tabular}{cccc}
    1 & 2 & 0 & 0 \\
    0 & 0 & 0 & 3 \\
    0 & 0 & 0 & 4
  \end{tabular}
\end{center}

The non-zero elements of this matrix are

\begin{eqnarray*}
   (1, 1)  & = & 1 \\
   (1, 2)  & = & 2 \\
   (2, 4)  & = & 3 \\
   (3, 4)  & = & 4
\end{eqnarray*}

This will be stored as three vectors $cidx$, $ridx$ and $data$,
representing the column indexing, row indexing and data
respectively. The contents of these three vectors for the above matrix
will be

\begin{eqnarray*}
  cidx & = & [0, 1, 2, 2, 4] \\
  ridx & = & [0, 0, 1, 2] \\
  data & = & [1, 2, 3, 4]
\end{eqnarray*}

Note that this is the representation of these elements with the first
row and column assumed to start at zero, while in {\Octave} itself the
row and column indexing starts at one. With the above representation,
the number of elements in the $i^{th}$ column is given by $cidx(i+1) -
cidx(i)$.

Although {\Octave} uses a compressed column format, it should be noted
that compressed row formats are equally possible. However,in the
context of mixed operations between mixed sparse and dense matrices,
it makes sense that the elements of the sparse matrices are in the
same order as the dense matrices. {\Octave} stores dense matrices in
column major ordering, and so sparse matrices are equally stored in
this manner.

A further constraint on the sparse matrix storage used by {\Octave} is that 
all elements in the rows are stored in increasing order of their row
index, which makes certain operations faster. However, it imposes
the need to sort the elements on the creation of sparse matrices. Having
unordered elements is potentially an advantage in that it makes operations
such as concatenating two sparse matrices together easier and faster, however
it adds complexity and speed problems elsewhere.

\subsection{Creating Sparse Matrices}
\label{sec:creation}

There are several means to create sparse matrices

\begin{itemize}
\item \textit{Returned from a function}:
There are many functions that directly return sparse matrices. These
include \textit{speye}, \textit{sprand}, \textit{spdiag}, etc.

\item \textit{Constructed from matrices or vectors}:
The function \textit{sparse} allows a sparse matrix to be constructed
from three vectors representing the row, column and
data. Alternatively, the function \textit{spconvert} uses a three
column matrix format to allow easy importation of data from elsewhere.

\item \textit{Created and then filled}:
The function \textit{sparse} or \textit{spalloc} can be used to create
an empty matrix that is then filled by the user

\item \textit{From a user binary program}:
The user can directly create the sparse matrix within an {\OctFile}.
\end{itemize}

There are several functions that return specific sparse
matrices. For example the sparse identity matrix is often
needed. It therefore has its own function to create it as
\textit{speye}$(n)$ or \textit{speye}$(r, c)$, which
creates an $n$-by-$n$ or $r$-by-$c$ sparse identity matrix.

Another typical sparse matrix that is often needed is a random
distribution of random elements. The functions \textit{sprand} and
\textit{sprandn} perform this for uniform and normal random
distributions of elements. They have exactly the same calling
convention, where \textit{sprand}$(r, c, d)$, creates
an $r$-by-$c$ sparse matrix with a density of filled elements
of $d$.

Other functions of interest that directly creates a sparse matrices, are
\textit{spdiag} or its generalization \textit{spdiags}, that can take 
the definition of the diagonals of the matrix and create the sparse matrix
that corresponds to this. For example

{\small
\begin{lstlisting}
s = spdiag (sparse(randn(1,n)),-1);
\end{lstlisting}
}

creates a sparse $(n+1)$-by-$(n+1)$ sparse matrix with a single
diagonal defined.

The recommended way for the user to create a sparse matrix, is to create 
two vectors containing the row and column index of the data and a third
vector of the same size containing the data to be stored. For example

{\small
\begin{lstlisting}
 function x = foo (r, j)
  idx = randperm (r);
  x = ([zeros(r-2,1); rand(2,1)])(idx);
 endfunction

 ri = [];
 ci = [];
 d = [];

 for j=1:c
    dtmp = foo (r, j);
    idx = find (dtmp != 0.);
    ri = [ri; idx];
    ci = [ci; j*ones(length(idx),1)]; 
    d = [d; dtmp(idx)];
 endfor
 s = sparse (ri, ci, d, r, c);
\end{lstlisting}
}

creates an $r$-by-$c$ sparse matrix with a random distribution of
2 elements per row. The elements of the vectors do not need to be sorted in
any particular order as {\Octave} will sort them prior to storing the
data. However, pre-sorting the data will make the creation of the sparse
matrix faster.

The function \textit{spconvert} takes a three or four column real matrix.
The first two columns represent the row and column index, respectively, and
the third and four columns, the real and imaginary parts of the sparse
matrix. The matrix can contain zero elements and the elements can be 
sorted in any order. Adding zero elements is a convenient way to define
the size of the sparse matrix. For example

{\small
\begin{lstlisting}
s = spconvert ([1 2 3 4; 1 3 4 4; 1 2 3 0]')
Compressed Column Sparse (rows=4, ...
  cols=4, nnz=3)
      (1 , 1) -> 1
      (2 , 3) -> 2
      (3 , 4) -> 3
\end{lstlisting}
}

An example of creating and filling a matrix might be

{\small
\begin{lstlisting}
k = 5;
nz = r * k;
s = spalloc (r, c, nz)
for j = 1:c
  idx = randperm (r);
  s (:, j) = [zeros(r - k, 1); ...
              rand(k, 1)] (idx);
endfor
\end{lstlisting}
}

It should be noted, that due to the way that the {\Octave} assignment
functions are written that the assignment will reallocate the memory
used by the sparse matrix at each iteration of the above loop.
Therefore the \textit{spalloc} function ignores the \textit{nz}
argument and does not preassign the memory for the matrix. Therefore,
code using the above structure should be vectorized to minimize the
number of assignments and reduce the number of memory allocations.

The above problem can be avoided in {\OctFiles}. However, the
construction of a sparse matrix from an {\OctFile} is more complex than
can be discussed in this brief introduction, and you are referred to
section~\ref{sec:octfiles}, to have a full description of the techniques
involved.

\subsection{Sparse Functions in {\Octave}}

An important consideration in the use of the sparse functions of
{\Octave} is that many of the internal functions of {\Octave}, such as
\textit{diag}, can not accept sparse matrices as an input. The sparse
implementation in {\Octave} therefore uses the \textit{dispatch}
function to overload the normal {\Octave} functions with equivalent
functions that work with sparse matrices. However, at any time the
sparse matrix specific version of the function can be used by
explicitly calling its function name. 

The table below lists all of the sparse functions of {\Octave}
together (with possible future extensions that are currently
unimplemented, listed last). Note that in this specific sparse forms
of the functions are typically the same as the general versions with a
\textit{sp} prefix. In the table below, and the rest of this article
the specific sparse versions of the functions are used.

\begin{itemize}
\item Generate sparse matrices:
  \textit{spalloc}, \textit{spdiags}, \textit{speye}, \textit{sp\-rand}, 
  \textit{sprandn}, (sprandsym)

\item Sparse matrix conversion:
  \textit{full}, \textit{sparse}, \textit{spconvert}, \textit{sp\-find}

\item Manipulate sparse matrices
  \textit{issparse}, \textit{nnz}, \textit{nonzeros}, \textit{nzmax},
  \textit{spfun}, \textit{spones}, \textit{spy},

\item Graph Theory:
  \textit{etree}, \textit{etreeplot}, \textit{gplot}, 
  \textit{treeplot}, (treelayout)

\item Sparse matrix reordering:
  \textit{ccolamd}, \textit{colamd}, \textit{colperm}, 
  \textit{csymamd}, \textit{symamd}, \textit{randperm}, (dmperm, symrcm)

\item Linear algebra:
  \textit{matrix\_type}, \textit{spchol}, \textit{spcholinv}, 
  \textit{spchol2inv}, \textit{spdet}, \textit{spinv}, \textit{spkron},
  \textit{splchol}, \textit{splu}, (condest, eigs, normest, 
  sprank, svds, spaugment, spqr)

\item Iterative techniques:
  \textit{luinc}, (bicg, bicgstab, cholinc, cgs, gmres, lsqr, minres, 
  pcg, pcr, qmr, symmlq)

\item Miscellaneous:
  \textit{spparms}, \textit{symbfact}, \textit{spstats}, 
  \textit{spprod}, \textit{spcumsum}, \textit{spsum},
  \textit{spsumsq}, \textit{spmin}, \textit{spmax}, \textit{spatan2}, 
  \textit{spdiag}
\end{itemize}

In addition all of the standard {\Octave} mapper functions (ie. basic
math functions that take a single argument) such as \textit{abs}, etc
can accept sparse matrices. The reader is referred to the documentation
supplied with these functions within {\Octave} itself for further
details.

\subsection{Sparse Return Types}
\label{sec:return}

The two basic reasons to use sparse matrices are to reduce the memory
usage and to not have to do calculations on zero elements. The two are
closely related and the computation time might be proportional to the
number of non-zero elements or a power of the number of non-zero
elements depending on the operator or function involved.

Therefore, there is a certain density of non-zero elements of a matrix 
where it no longer makes sense to store it as a sparse matrix, but rather
as a full matrix. For this reason operators and functions that have a 
high probability of returning a full matrix will always return one. For
example adding a scalar constant to a sparse matrix will almost always
make it a full matrix, and so the example

{\small
\begin{lstlisting}
speye(3) + 0
  1  0  0
  0  1  0
  0  0  1
\end{lstlisting}
}

returns a full matrix as can be seen. Additionally all sparse functions
test the amount of memory occupied by the sparse matrix to see if the 
amount of storage used is larger than the amount used by the full 
equivalent. Therefore \textit{speye(2) * 1} will return a full matrix as
the memory used is smaller for the full version than the sparse version.

As all of the mixed operators and functions between full and sparse 
matrices exist, in general this does not cause any problems. However,
one area where it does cause a problem is where a sparse matrix is
promoted to a full matrix, where subsequent operations would re-sparsify
the matrix. Such cases are rare, but can be artificially created, for
example \textit{(fliplr(speye(3)) + speye(3)) - speye(3)} 
gives a full matrix when it should give a sparse one. In general, where 
such cases  occur, they impose only a small memory penalty.

There is however one known case where this behavior of {\Octave}'s
sparse matrices will cause a problem. That is in the handling of the
\textit{spdiag} function. Whether \textit{spdiag} returns a sparse or 
full matrix depends on the type of its input arguments. So 

{\small
\begin{lstlisting}
 a = diag (sparse([1,2,3]), -1);
\end{lstlisting}
}

should return a sparse matrix. To ensure this actually happens, the
\textit{sparse} function, and other functions based on it like 
\textit{speye}, always returns a sparse matrix, even if the memory 
used will be larger than its full representation.

\subsection{Finding out Information about Sparse Matrices}
\label{sec:info}

There are a number of functions that allow information concerning
sparse matrices to be obtained. The most basic of these is
\textit{issparse} that identifies whether a particular {\Octave}
object is in fact a sparse matrix. 

Another very basic function is \textit{nnz} that returns the number of
non-zero entries there are in a sparse matrix, while the function
\textit{nzmax} returns the amount of storage allocated to the sparse
matrix. Note that {\Octave} tends to crop unused memory at the first
opportunity for sparse objects. There are some cases of user created
sparse objects where the value returned by \textit{nzmaz} will not be
the same as \textit{nnz}, but in general they will give the same
result. The function \textit{spstats} returns some basic statistics on
the columns of a sparse matrix including the number of elements, the
mean and the variance of each column.

When solving linear equations involving sparse matrices {\Octave}
determines the means to solve the equation based on the type of the
matrix as discussed in section~\ref{sec:linalg}. {\Octave} probes
the matrix type when the div ($/$) or ldiv ($\backslash$) operator
is first used with the matrix and then caches the type. However the
\textit{matrix\_type} function can be used to determine the type of
the sparse matrix prior to use of the div or ldiv operators. For example

{\small
\begin{lstlisting}
a = tril (sprandn(1024, 1024, 0.02), -1) ...
    + speye(1024); 
matrix_type (a);
ans = Lower
\end{lstlisting}
}

show that {\Octave} correctly determines the matrix type for lower
triangular matrices. \textit{matrix\_type} can also be used to force
the type of a matrix to be a particular type. For example

{\small
\begin{lstlisting}
a = matrix_type (tril (sprandn (1024, ...
   1024, 0.02), -1) + speye(1024), 'Lower');
\end{lstlisting}
}

This allows the cost of determining the matrix type to be
avoided. However, incorrectly defining the matrix type will result in
incorrect results from solutions of linear equations, and so it is
entirely the responsibility of the user to correctly identify the
matrix type

There are several graphical means of finding out information about
sparse matrices. The first is the \textit{spy} command, which displays
the structure of the non-zero elements of the matrix, as can be seen
in Figure~\ref{fig:eidorsmatrix}. More advanced graphical information
can be obtained with the \textit{treeplot}, \textit{etreeplot} and
\textit{gplot} commands.

One use of sparse matrices is in graph theory, where the
interconnections between nodes is represented as an adjacency
matrix~\cite{bondy72}. That is, if the $i$-th node in a graph is
connected to the $j$-th node. Then the $ij$-th node (and in the case
of undirected graphs the $ji$-th node) of the sparse adjacency matrix
is non-zero. If each node is then associated with a set of
co-ordinates, then the \textit{gplot} command can be used to
graphically display the interconnections between nodes.

As a trivial example of the use of \textit{gplot}, consider the
example

\begin{figure}
  \begin{minipage}[b]{1.0\linewidth}
    \centering
    \centerline{\epsfig{figure=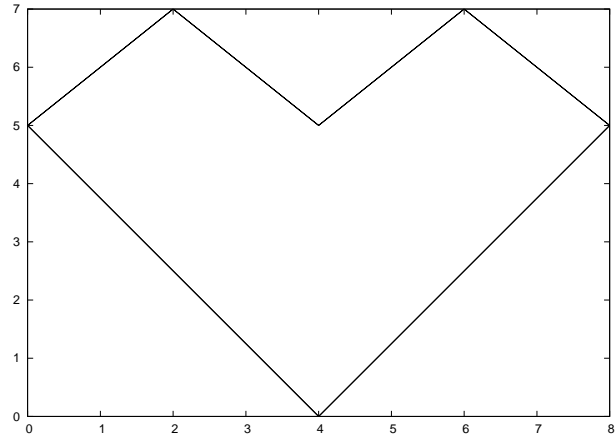,width=8.5cm}}
  \end{minipage}
  \begin{center}
  \caption{Simple use of the \textit{gplot} command as discussed in Section~\ref{sec:info}.}
  \label{fig:gplot}
  \end{center}
\end{figure}

{\small
\begin{lstlisting}
A = sparse([2,6,1,3,2,4,3,5,4,6,1,5],
    [1,1,2,2,3,3,4,4,5,5,6,6],1,6,6);
xy = [0,4,8,6,4,2;5,0,5,7,5,7]';
gplot(A,xy)
\end{lstlisting}
}

which creates an adjacency matrix \textit{A} where node 1 is connected
to nodes 2 and 6, node 2 with nodes 1 and 3, etc. The co-ordinates of
the nodes is given in the \textit{n}-by-\textit{2} matrix \textit{xy}.
The output of the \textit{gplot} command can be seen in
Figure~\ref{fig:gplot}

The dependences between the nodes of a Cholesky factorization can be
calculated in linear time without explicitly needing to calculate the
Cholesky factorization by the \textit{etree} command. This command
returns the elimination tree of the matrix and can be displayed
grapically by the command \textit{treeplot(etree(A))} if
\textit{A} is symmetric or \textit{treeplot(etree(A+A'))}
otherwise.

\subsection{Mathematical Considerations}
\label{sec:consider}

The attempt has been made to make sparse matrices behave in exactly the
same manner as their full counterparts. However, there are certain differences
between full and sparse behavior and with the sparse implementations
in other software tools.

Firstly, the \textit{./} and \textit{.$^\wedge$} operators must be used with 
care. Consider what the examples

\begin{eqnarray*}
  s & = & speye (4); \\
  a1 & = & s .^\wedge 2; \\
  a2 & = & s .^\wedge s; \\
  a3 & = & s .^\wedge -2; \\
  a4 & = & s ./ 2; \\
  a5 & = & 2 ./ s; \\
  a6 & = & s ./ s;
\end{eqnarray*}

will give. The first example of $s$ raised to the power of 2 causes
no problems. However $s$ raised element-wise to itself involves a
a large number of terms \textit{0 .$^\wedge$ 0} which is 1. Therefore 
\textit{s .$^\wedge$ s} is a full matrix.

Likewise \textit{s .$^\wedge$ -2} involves terms terms like \textit{0
.$^\wedge$ -2} which is infinity, and so \textit{s .$^\wedge$ -2} is
equally a full matrix.

For the \textit{./} operator \textit{s ./ 2} has no problems, but 
\textit{2 ./ s} involves a large number of infinity terms as well
and is equally a full matrix. The case of \textit{s ./ s}
involves terms like \textit{0 ./ 0} which is a \textit{NaN} and so this
is equally a full matrix with the zero elements of $s$ filled with
\textit{NaN} values. The above behavior is consistent with full 
matrices, but is not consistent with sparse implementations in
{\Matlab}~\cite{gilbert92sparse}. If the user requires the same 
behavior as in {\Matlab} then for example for the case of 
\textit{2 ./ s} then appropriate code is

{\small
\begin{lstlisting}
function z = f(x), z = 2 ./ x; endfunction
spfun (@f, s);
\end{lstlisting}
}

and the other examples above can be implemented similarly.

A particular problem of sparse matrices comes about due to the fact
that as the zeros are not stored, the sign-bit of these zeros is
equally not stored. In certain cases the sign-bit of zero is
important~\cite{Kahan.1987}. For example

{\small
\begin{lstlisting}
 a = 0 ./ [-1, 1; 1, -1];
 b = 1 ./ a
    -Inf            Inf
     Inf           -Inf
 c = 1 ./ sparse (a)
     Inf            Inf
     Inf            Inf
\end{lstlisting}
}
 
To correct this behavior would mean that zero elements with a negative
sign-bit would need to be stored in the matrix to ensure that their 
sign-bit was respected. This is not done at this time, for reasons of
efficiency, and so the user is warned that calculations where the sign-bit
of zero is important must not be done using sparse matrices.

In general any function or operator used on a sparse matrix will
result in a sparse matrix with the same or a larger number of non-zero
elements than the original matrix. This is particularly true for the
important case of sparse matrix factorizations. The usual way to
address this is to reorder the matrix, such that its factorization is
sparser than the factorization of the original matrix. That is the
factorization of $L U = P S Q$ has sparser terms $L$ and $U$ than the
equivalent factorization $L U = S$.

Several functions are available to reorder depending on the type of the
matrix to be factorized. If the matrix is symmetric positive-definite,
then \textit{symamd} or \textit{csymamd} should be used. Otherwise
\textit{colamd} or \textit{ccolamd} should be used. For completeness
the reordering functions \textit{colperm} and \textit{randperm} are
also available.

\begin{figure}
  \begin{minipage}[b]{1.0\linewidth}
    \centering
    \centerline{\epsfig{figure=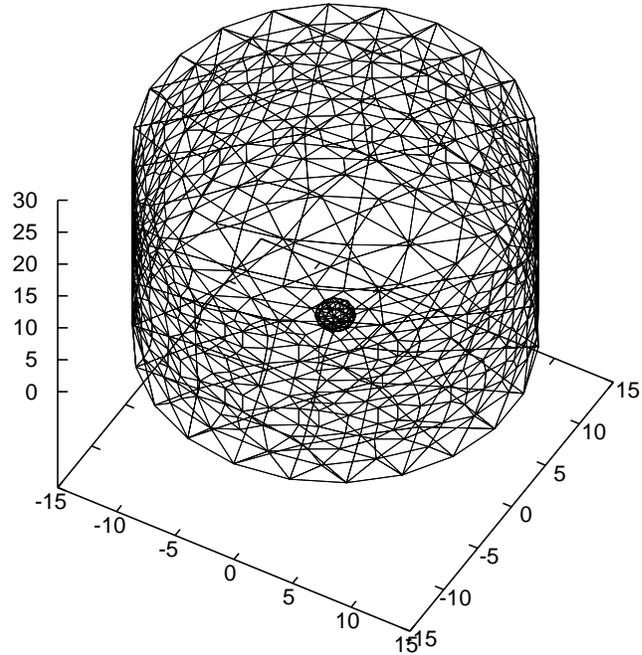,width=8.5cm}}
  \end{minipage}
  \begin{center}
  \caption{Geometry of FEM model of phantom ball model from EIDORS project~\cite{eidors.website,eidors2006}}
  \label{fig:eidorsball}
  \end{center}
\end{figure}

\begin{figure}
  \begin{minipage}[b]{1.0\linewidth}
    \centering
    \centerline{\epsfig{figure=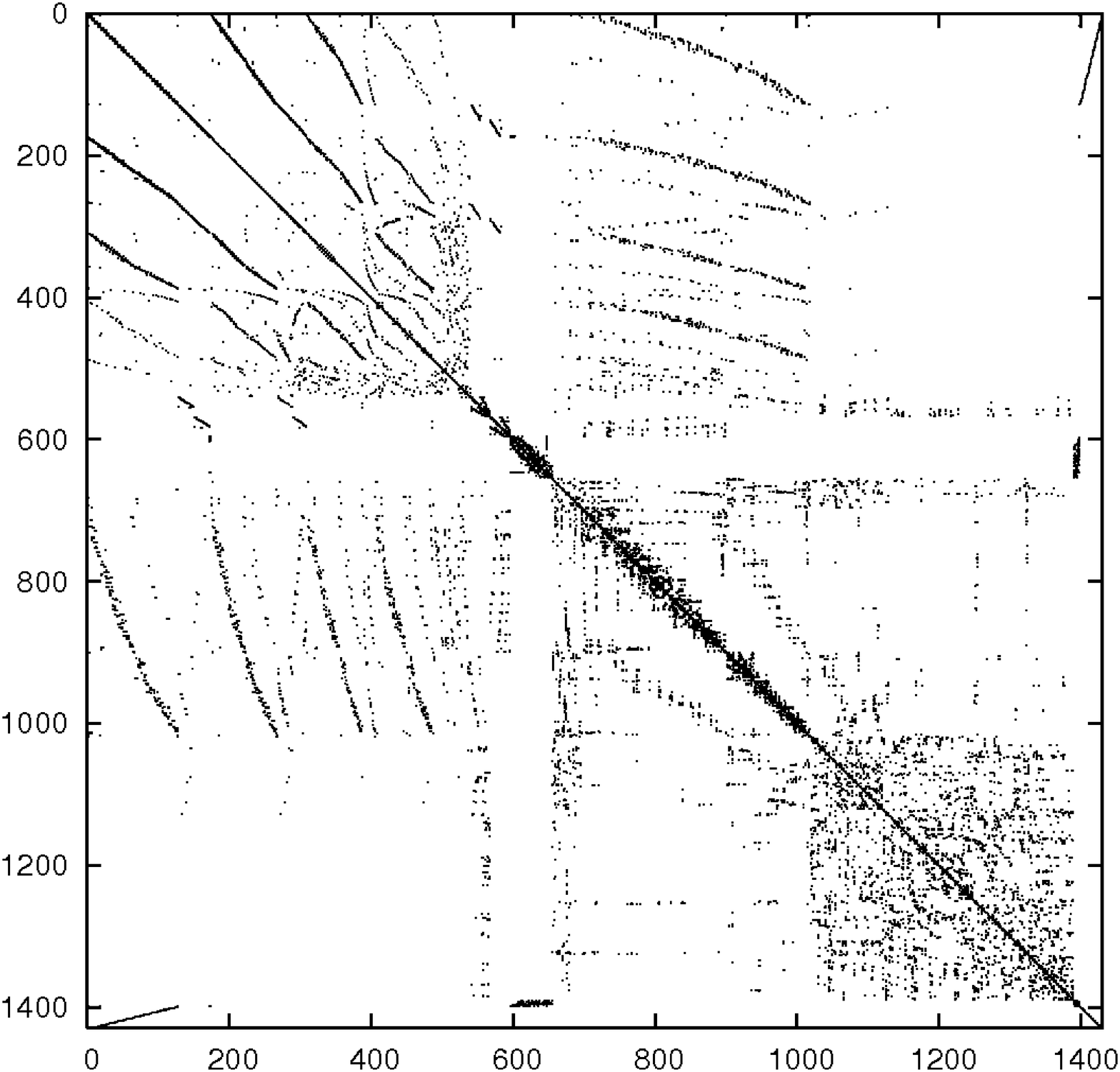,width=8.5cm}}
  \end{minipage}
  \begin{center}
  \caption{Structure of the sparse matrix derived from EIDORS phantom ball model~\cite{eidors.website,eidors2006}}
  \label{fig:eidorsmatrix}
  \end{center}
\end{figure}

As an example, consider the ball model which is given as an
example in the EIDORS project~\cite{eidors.website,eidors2006}, as shown in
Figure~\ref{fig:eidorsball}. The structure of the original matrix
derived from this problem can be seen with the command
\textit{spy(A)}, as seen in Figure~\ref{fig:eidorsmatrix}.

\begin{figure}
  \begin{minipage}[b]{1.0\linewidth}
    \centering
    \centerline{\epsfig{figure=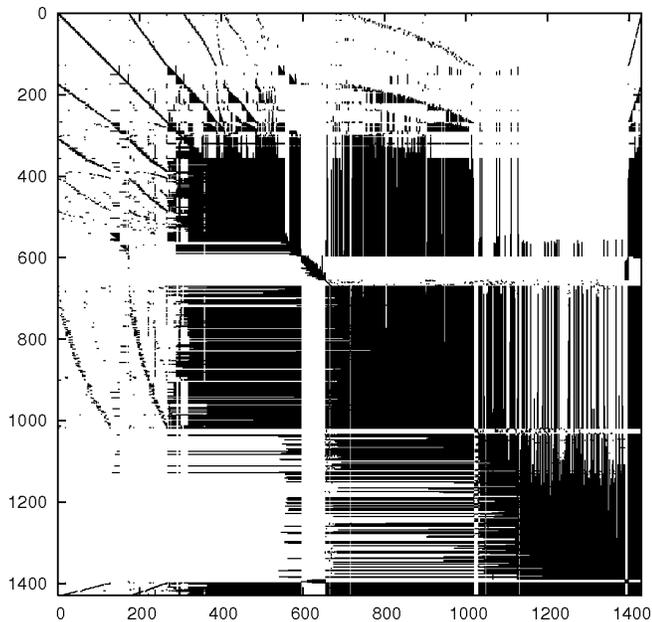,width=8.5cm}}
  \end{minipage}
  \begin{center}
  \caption{Structure of the un-permuted LU factorization of EIDORS ball problem}
  \label{fig:eidorslu}
  \end{center}
\end{figure}

The standard LU factorization of this matrix, with row pivoting can be
obtained by the same command that would be used for a full
matrix. This can be visualized with the command \textit{[l, u, p] =
lu(A); spy(l+u);} as seen in Figure~\ref{fig:eidorslu}. The original
matrix had 17825 non-zero terms, while this LU factorization has
531544 non-zero terms, which is a significant level of fill in of the
factorization and represents a large overhead in working with this
matrix.

The appropriate sparsity preserving permutation of the original
matrix is given by \textit{colamd} and the factorization using this
reordering can be visualized using the command \textit{q = colamd(A);
[l, u, p] = lu(A(:,q)); spy(l+u)}. This gives 212044 non-zero terms
which is a significant improvement.

\begin{figure}
  \begin{minipage}[b]{1.0\linewidth}
    \centering
    \centerline{\epsfig{figure=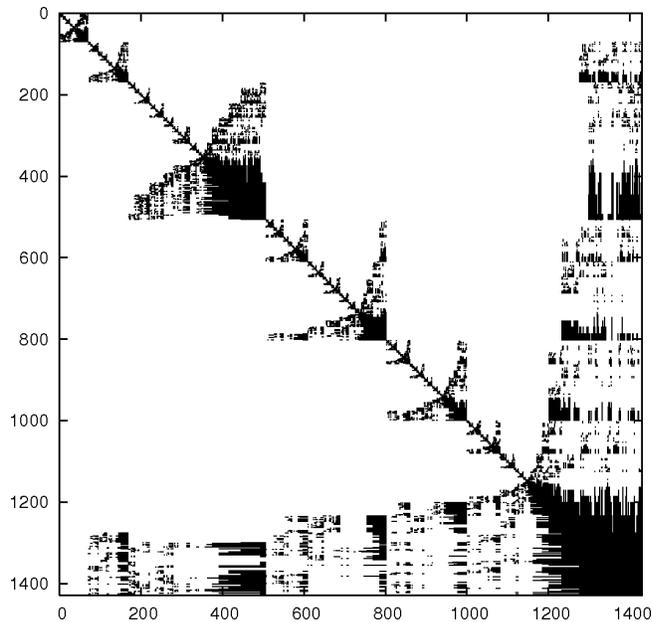,width=8.5cm}}
  \end{minipage}
  \begin{center}
  \caption{Structure of the permuted LU factorization of EIDORS ball problem}
  \label{fig:eidorspermlu}
  \end{center}
\end{figure}

Furthermore, the underlying factorization software updates its
estimate of the optimal sparsity preserving reordering of the matrix
during the factorization, so can return an even sparser factorization.
In the case of the LU factorization this might be obtained with a
fourth return argument as \textit{[l, u, p, q] = lu(A);
spy(l+u)}. This factorization has 143491 non-zero terms, and its
structure can be seen in Figure~\ref{fig:eidorspermlu}.

Finally, {\Octave} implicitly reorders the matrix when using the div
($/$) and ldiv ($\backslash$) operators, and so no the user does not
need to explicitly reorder the matrix to maximize performance.

\section{Linear Algebra on Sparse Matrices}
\label{sec:linalg}

{\Octave} includes a polymorphic solver for sparse matrices, where
the exact solver used to factorize the matrix, depends on the
properties of the sparse matrix itself. Generally, the cost of
determining the matrix type is small relative to the cost of
factorizing the matrix itself, but in any case the matrix type is
cached once it is calculated, so that it is not re-determined each
time it is used in a linear equation.

Linear equations are solved using the following selection tree

\begin{enumerate}
\item If the matrix is not square go to 9.

\item If the matrix is diagonal, solve directly and go to 9

\item If the matrix is a permuted diagonal, solve directly taking into
account the permutations. Go to 9

\item If the matrix is banded and if the band density is less than that
given by \textit{spparms ("bandden")} continue, else go to 5.

\begin{enumerate}
\item If the matrix is tridiagonal and the right-hand side is not sparse 
continue, else go to 4b.

\begin{enumerate}
\item If the matrix is hermitian, with a positive real diagonal, attempt
      Cholesky factorization using \textit{Lapack} xPTSV.

\item If the above failed, or the matrix is not hermitian,
      use Gaussian elimination with pivoting using 
      \textit{Lapack} xGTSV, and go to 9.
\end{enumerate}

\item If the matrix is hermitian with a positive real diagonal, attempt
      a Cholesky factorization using \textit{Lapack} xPBTRF.

\item if the above failed or the matrix is not hermitian with a positive
      real diagonal use Gaussian elimination with pivoting using 
      \textit{Lapack} xGBTRF, and go to 9.
\end{enumerate}

\item If the matrix is upper or lower triangular perform a sparse forward
or backward substitution, and go to 9

\item If the matrix is a upper triangular matrix with column permutations
or lower triangular matrix with row permutations, perform a sparse forward 
or backward substitution, and go to 9

\item If the matrix is hermitian with a real positive diagonal, attempt
a sparse Cholesky factorization using CHOLMOD~\cite{davis.2006}.

\item If the sparse Cholesky factorization failed or the matrix is not
hermitian, perform LU factorization using UMFPACK~\cite{davis.2004}.

\item If the matrix is not square, or any of the previous solvers flags
a singular or near singular matrix, find a minimum norm solution.

\end{enumerate}

The band density is defined as the number of non-zero values in the band
divided by the number of values in the band. The banded matrix
solvers can be entirely disabled by using \textit{spparms} to set 
\textit{bandden} to 1 (i.e. \textit{spparms ("bandden", 1)}).

The QR solver factorizes the problem with a
Dulmage-Mendhelsohn~\cite{ashcraft96applications}, to seperate the
problem into blocks that can be treated as over-determined, multiple
well determined blocks, and a final over-determined block. For
matrices with blocks of strongly connectted nodes this is a big win
as LU decomposition can be used for many blocks. It also significantly
improves the chance of finding a solution to ill-conditioned problems
rather than just returning a vector of \textit{NaN}'s.

All of the solvers above, can calculate an estimate of the condition
number. This can be used to detect numerical stability problems in the
solution and force a minimum norm solution to be used. However, for
narrow banded, triangular or diagonal matrices, the cost of
calculating the condition number is significant, and can in fact
exceed the cost of factoring the matrix. Therefore the condition
number is not calculated in these case, and octave relies on simplier
techniques to detect sinular matrices or the underlying LAPACK code in
the case of banded matrices.

The user can force the type of the matrix with the \textit{matrix\_type}
function. This overcomes the cost of discovering the type of the matrix.
However, it should be noted incorrectly identifying the type of the matrix
will lead to unpredictable results, and so \textit{matrix\_type} should be
used with care.

\section{Benchmarking of Octave Sparse Matrix Implementation}
\label{sec:bench}

It is a truism that all benchmarks should be treated with care. The
speed of a software package is determined by a large number of
factors, including the particular problem treated and the
configuration of the machine on which the benchmarks were run.
Therefore the benchmarks presented here should be treated as
indicative of the speed a user might expect.

That being said we attempt to examine the speed of several fundamental
operators for use with sparse matrices. These being the addition (+),
multiplication (*) and left-devision ($\backslash$) operators. The
basic test code used to perform these tests is given by

{\small
\begin{lstlisting}
time = 0;
n = 0;
while (time < tmin || n < nrun)
  clear a, b;
  a = sprand (order, order, density);
  t = cputime ();
  b = a OP a;
  time = time + cputime () - t;
  n = n + 1;
end
time = time / n;
\end{lstlisting}
}

where \textit{nrun} was 5, \textit{tmin} was 1 second and \textit{OP}
was either +, or *. The left-division operator poses particular problems
for benchmarking that will be discussed later.

Although the \textit{cputime} function only has a resolution of 0.01
seconds, running the command multiple times and limited by the minimum
run time of \textit{tmin} seconds allows this precision to be extended.
Running the above code for various matrix orders and densities results 
in the summary of execution times as seen in Table~\ref{tab:bench1}.

\begin{table}
{\small
\begin{tabular}{|c|c||c|c|c|c|}\hline
 Order & Den-     & \multicolumn{4}{|c|}{Execution Time for Operator (sec)} \\ \cline{3-6}
       &  sity    & \multicolumn{2}{|c|}{\Matlab} & \multicolumn{2}{|c|}{\Octave} \\ \cline{3-6}
       &          &     +     &     *     &     +     &     *     \\ \hline
  500  &  1e-02   &  0.00049  &  0.00250  &  0.00039  &  0.00170  \\
  500  &  1e-03   &  0.00008  &  0.00009  &  0.00022  &  0.00026  \\
  500  &  1e-04   &  0.00005  &  0.00007  &  0.00020  &  0.00024  \\
  500  &  1e-05   &  0.00004  &  0.00007  &  0.00021  &  0.00015  \\
  500  &  1e-06   &  0.00006  &  0.00007  &  0.00020  &  0.00021  \\
 1000  &  1e-02   &  0.00179  &  0.02273  &  0.00092  &  0.00990  \\
 1000  &  1e-03   &  0.00021  &  0.00027  &  0.00029  &  0.00042  \\
 1000  &  1e-04   &  0.00011  &  0.00013  &  0.00023  &  0.00026  \\
 1000  &  1e-05   &  0.00012  &  0.00011  &  0.00028  &  0.00023  \\
 1000  &  1e-06   &  0.00012  &  0.00010  &  0.00021  &  0.00022  \\
 2000  &  1e-02   &  0.00714  &  0.23000  &  0.00412  &  0.07049  \\
 2000  &  1e-03   &  0.00058  &  0.00165  &  0.00055  &  0.00135  \\
 2000  &  1e-04   &  0.00032  &  0.00026  &  0.00026  &  0.00033  \\
 2000  &  1e-05   &  0.00019  &  0.00020  &  0.00022  &  0.00026  \\
 2000  &  1e-06   &  0.00018  &  0.00018  &  0.00024  &  0.00023  \\
 5000  &  1e-02   &  0.05100  &  3.63200  &  0.02652  &  0.95326  \\
 5000  &  1e-03   &  0.00526  &  0.03000  &  0.00257  &  0.01896  \\
 5000  &  1e-04   &  0.00076  &  0.00083  &  0.00049  &  0.00074  \\
 5000  &  1e-05   &  0.00051  &  0.00051  &  0.00031  &  0.00043  \\
 5000  &  1e-06   &  0.00048  &  0.00055  &  0.00028  &  0.00026  \\
10000  &  1e-02   &  0.22200  &  24.2700  &  0.10878  &  6.55060  \\
10000  &  1e-03   &  0.02000  &  0.30000  &  0.01022  &  0.18597  \\
10000  &  1e-04   &  0.00201  &  0.00269  &  0.00120  &  0.00252  \\
10000  &  1e-05   &  0.00094  &  0.00094  &  0.00047  &  0.00074  \\
10000  &  1e-06   &  0.00110  &  0.00098  &  0.00039  &  0.00055  \\
20000  &  1e-03   &  0.08286  &  2.65000  &  0.04374  &  1.71874  \\
20000  &  1e-04   &  0.00944  &  0.01923  &  0.00490  &  0.01500  \\
20000  &  1e-05   &  0.00250  &  0.00258  &  0.00092  &  0.00149  \\
20000  &  1e-06   &  0.00189  &  0.00161  &  0.00058  &  0.00121  \\
50000  &  1e-04   &  0.05500  &  0.39400  &  0.02794  &  0.28076  \\
50000  &  1e-05   &  0.00823  &  0.00877  &  0.00406  &  0.00767  \\
50000  &  1e-06   &  0.00543  &  0.00610  &  0.00154  &  0.00332  \\ \hline
\end{tabular}
}
\caption{Benchmark of basic operators on {\Matlab} R14sp2 against {\Octave} 2.9.5, on a Pentium 4M 1.6GHz machine with 1GB of memory.\label{tab:bench1}}
\end{table}

The results for the small low density problems in
Table~\ref{tab:bench1} are interesting (cf. Matrix order of 500, with
densities lower than 1e-03), as they seem to indicate that there is a
small incompressible execution time for both {\Matlab} and
{\Octave}. This is probably due to the overhead associated with the
parsing of the language and the calling of the underlying function
responsible for the operator. On the test machine this time was
approximately 200 $\mu$s for {\Octave} for both operators, while for
{\Matlab} this appears to be 70 and 40 $\mu$s for the * and +
operators respectively. So in this class of problems {\Matlab}
outperforms {\Octave} for both operators.  However, when the matrix
order or density increases it can be seen that {\Octave} significantly
out-performs {\Matlab} for both operators.

When considering the left-division operator, we can not use randomly
created matrices. The reason is that the fill-in, or rather the
potential to reduce the fill-in with appropriate matrix re-ordering,
during matrix factorization is determined by the structure of the
matrix imposed by the problem it represents. As random matrices have
no structure, factorization of random matrices results in extremely
large levels of matrix fill-in, even with matrix re-ordering. 
Therefore, to benchmark the left-division ($\backslash$) operator, 
we have selected a number of test matrices that are publicly 
available~\cite{davis.1997}, and modify the benchmark code as

{\small
\begin{lstlisting}
time = 0;
n = 0;
while (time < tmin || n < nrun)
  clear a, b;
  load test.mat % Get matrix 'a'
  x = ones(order,1);
  t = cputime ();
  b = a \ x;
  time = time + cputime () - t;
  n = n + 1;
end
time = time / n;
\end{lstlisting}
}

All the the matrices in the University of Florida Sparse
Matrix~\cite{davis.1997} that met the following criteria were used

\begin{itemize}
\item Has real or complex data available, and not just the structure,
\item Has between 10,000 and 1,000,000 non-zero element,
\item Has equal number of rows and columns,
\item The solution did not require more than 1GB of memory, to avoid issues
with memory.
\end{itemize}

When comparing the benchmarks for the left-division operator it must
be considered that the matrices in the collection used represent an
arbitrary sampling of the available sparse matrix problems. It is
therefore difficult to treat the data in aggregate, and so we present
the raw data below so that the reader might compare the benchmark for
a particular matrix class that interests them.

The performance of the {\Matlab} and {\Octave} left-division operators
is affected by the \textit{spparms} function. In particular the density
of terms in a banded matrix that is needed to force the solver to use the
LAPACK banded solvers rather than the generic solvers is determined by the
command \textit{spparms('bandden',val)}. The default density of 0.5
was used for both {\Matlab} and {\Octave}.

Five classes of problems were represented in the matrices treated. These
are

\begin{itemize}
\item Banded positive definite and factorized with the LAPACK xPBTRF function,
\item General banded matrix and factorized with the LAPACK xGBTRF function,
\item Positive definite and treated by the Cholesky solvers of {\Matlab} 
and {\Octave},
\item Sparse LU decomposition with UMFPACK, and
\item Singular matrices that were treated via QR decomposition.
\end{itemize}
 
Also, it should be noted that the LAPACK solvers, and dense BLAS
kernels of the UMFPACK and CHOLMOD solvers were accelerated using the
ATLAS~\cite{whaley97automatically} versions of the LAPACK and BLAS
functions. The exact manner in which the ATLAS library is compiled
might have an affect on the performance, and therefore the benchmarks
might measure the relative performance of the different versions of
ATLAS rather than the performance of {\Octave} and {\Matlab}. To avoid
this issue {\Octave} was forced to use the {\Matlab} ATLAS libraries
with the use of the Unix LD\_PRELOAD command.

\begin{table*}
\begin{center}
{\tiny
\begin{tabular}{|c|c|c|c||c|c||c|c|c|c||c|c|}\hline
                 Matrix        & Order &   NNZ  & S$^\dag$ & \multicolumn{2}{|c|}{Execution Time} &
                 Matrix        & Order &   NNZ  & S$^\dag$ & \multicolumn{2}{|c|}{Execution Time} \\
                               &       &        &    &  \multicolumn{2}{|c|}{for Operator (sec)} &
                               &       &        &    &  \multicolumn{2}{|c|}{for Operator (sec)} \\ \cline{5-6} \cline{11-12}
                               &       &        &    & Matlab    &   Octave &
                               &       &        &    & Matlab    &   Octave \\ \hline
                    Bai/dw1024 &  2048 &  10114 &  8 &   0.05000 &   0.03585 &                        HB/nos3 &   960 &  15844 &  7 &   0.04417 &   0.01050 \\
               Boeing/bcsstm38 &  8032 &  10485 &  9 &   0.04333 &   0.02490 &                    Bai/rbsa480 &   480 &  17088 &  8 &   0.05000 &   0.02905 \\
                  Zitney/extr1 &  2837 &  10967 &  8 &   0.03846 &   0.02052 &                    Bai/rbsb480 &   480 &  17088 &  8 &   0.04545 &   0.02575 \\
             vanHeukelum/cage8 &  1015 &  11003 &  8 &   0.07714 &   0.05039 &             Hollinger/g7jac010 &  2880 &  18229 &  8 &   0.18000 &   0.13538 \\
                    FIDAP/ex32 &  1159 &  11047 &  8 &   0.04333 &   0.02354 &           Hollinger/g7jac010sc &  2880 &  18229 &  8 &   0.15600 &   0.13778 \\
        Sandia/adder\_dcop\_05 &  1813 &  11097 &  8 &   0.03000 &   0.01693 &                   Mallya/lhr01 &  1477 &  18427 &  8 &   0.05667 &   0.02982 \\
        Sandia/adder\_dcop\_04 &  1813 &  11107 &  8 &   0.02889 &   0.01680 &                    HB/bcsstk09 &  1083 &  18437 &  7 &   0.05778 &   0.02012 \\
        Sandia/adder\_dcop\_03 &  1813 &  11148 &  8 &   0.03059 &   0.01690 &                     FIDAP/ex21 &   656 &  18964 &  8 &   0.03846 &   0.02390 \\
        Sandia/adder\_dcop\_01 &  1813 &  11156 &  8 &   0.02941 &   0.01670 &                     Wang/wang1 &  2903 &  19093 &  8 &   0.23400 &   0.14818 \\
           Sandia/init\_adder1 &  1813 &  11156 &  8 &   0.02889 &   0.01667 &                     Wang/wang2 &  2903 &  19093 &  8 &   0.22800 &   0.14798 \\
        Sandia/adder\_dcop\_06 &  1813 &  11224 &  8 &   0.02833 &   0.01693 &               Brethour/coater1 &  1348 &  19457 &  9 &   0.19000 &   0.07413 \\
        Sandia/adder\_dcop\_07 &  1813 &  11226 &  8 &   0.02889 &   0.01670 &                    HB/bcsstm12 &  1473 &  19659 &  7 &   0.07000 &   0.01037 \\
        Sandia/adder\_dcop\_10 &  1813 &  11232 &  8 &   0.03000 &   0.01670 &                     Hamm/add32 &  4960 &  19848 &  8 &   0.06500 &   0.03869 \\
        Sandia/adder\_dcop\_09 &  1813 &  11239 &  8 &   0.02889 &   0.01706 &                    Bai/olm5000 &  5000 &  19996 & 4d &   0.00463 &   0.00546 \\
        Sandia/adder\_dcop\_08 &  1813 &  11242 &  8 &   0.03118 &   0.01696 &                       Gset/G57 &  5000 &  20000 &  8 &   0.15800 &   0.09332 \\
        Sandia/adder\_dcop\_11 &  1813 &  11243 &  8 &   0.02889 &   0.01693 &                    HB/sherman3 &  5005 &  20033 &  8 &   0.12200 &   0.07285 \\
        Sandia/adder\_dcop\_13 &  1813 &  11245 &  8 &   0.02889 &   0.01751 &                    Shyy/shyy41 &  4720 &  20042 &  8 &   0.08833 &   0.05149 \\
        Sandia/adder\_dcop\_19 &  1813 &  11245 &  8 &   0.02889 &   0.01693 &                     Bai/rw5151 &  5151 &  20199 &  8 &   0.21600 &   0.13078 \\
        Sandia/adder\_dcop\_44 &  1813 &  11245 &  8 &   0.02889 &   0.01713 &            Oberwolfach/t3dl\_e & 20360 &  20360 & 4c &   0.00105 &   0.00327 \\
        Sandia/adder\_dcop\_02 &  1813 &  11246 &  8 &   0.03059 &   0.01769 &                Boeing/bcsstm35 & 30237 &  20619 &  9 &   0.09333 &   0.05429 \\
        Sandia/adder\_dcop\_12 &  1813 &  11246 &  8 &   0.02889 &   0.01606 &                  Grund/bayer08 &  3008 &  20698 &  8 &   0.09667 &   0.05766 \\
        Sandia/adder\_dcop\_14 &  1813 &  11246 &  8 &   0.02889 &   0.01683 &                  Grund/bayer05 &  3268 &  20712 &  8 &   0.02125 &   0.00998 \\
        Sandia/adder\_dcop\_15 &  1813 &  11246 &  8 &   0.02833 &   0.01676 &                  Grund/bayer06 &  3008 &  20715 &  8 &   0.10000 &   0.05799 \\
        Sandia/adder\_dcop\_16 &  1813 &  11246 &  8 &   0.02778 &   0.01683 &                    HB/sherman5 &  3312 &  20793 &  8 &   0.09333 &   0.05259 \\
        Sandia/adder\_dcop\_17 &  1813 &  11246 &  8 &   0.02889 &   0.01727 &                    Wang/swang1 &  3169 &  20841 &  8 &   0.08500 &   0.04817 \\
        Sandia/adder\_dcop\_18 &  1813 &  11246 &  8 &   0.02889 &   0.01703 &                    Wang/swang2 &  3169 &  20841 &  8 &   0.08500 &   0.04808 \\
        Sandia/adder\_dcop\_20 &  1813 &  11246 &  8 &   0.02889 &   0.01680 &                  Grund/bayer07 &  3268 &  20963 &  8 &   0.01923 &   0.01010 \\
        Sandia/adder\_dcop\_21 &  1813 &  11246 &  8 &   0.02778 &   0.01686 &                    HB/bcsstm13 &  2003 &  21181 &  9 &   0.10200 &   0.06949 \\
        Sandia/adder\_dcop\_22 &  1813 &  11246 &  8 &   0.03000 &   0.01700 &              Bomhof/circuit\_2 &  4510 &  21199 &  8 &   0.04250 &   0.02395 \\
        Sandia/adder\_dcop\_23 &  1813 &  11246 &  8 &   0.03059 &   0.01690 &                Boeing/bcsstk34 &   588 &  21418 &  7 &   0.09333 &   0.01539 \\
        Sandia/adder\_dcop\_24 &  1813 &  11246 &  8 &   0.02889 &   0.01727 &               TOKAMAK/utm1700b &  1700 &  21509 &  8 &   0.08143 &   0.05009 \\
        Sandia/adder\_dcop\_25 &  1813 &  11246 &  8 &   0.03000 &   0.01693 &                    HB/bcsstk10 &  1086 &  22070 &  7 &   0.11800 &   0.00826 \\
        Sandia/adder\_dcop\_26 &  1813 &  11246 &  8 &   0.03000 &   0.01700 &                    HB/bcsstm10 &  1086 &  22092 &  8 &   0.14000 &   0.02008 \\
        Sandia/adder\_dcop\_27 &  1813 &  11246 &  8 &   0.02889 &   0.01713 &                 Hamrle/Hamrle2 &  5952 &  22162 &  8 &   0.11000 &   0.06299 \\
        Sandia/adder\_dcop\_28 &  1813 &  11246 &  8 &   0.02889 &   0.01680 &                     FIDAP/ex33 &  1733 &  22189 &  7 &   0.06875 &   0.01205 \\
        Sandia/adder\_dcop\_29 &  1813 &  11246 &  8 &   0.02789 &   0.01680 &                      HB/saylr4 &  3564 &  22316 &  8 &   0.18000 &   0.11338 \\
        Sandia/adder\_dcop\_30 &  1813 &  11246 &  8 &   0.02889 &   0.01680 &                     FIDAP/ex22 &   839 &  22460 &  8 &   0.04154 &   0.02234 \\
        Sandia/adder\_dcop\_31 &  1813 &  11246 &  8 &   0.02889 &   0.01693 &                   Zitney/hydr1 &  5308 &  22680 &  8 &   0.09000 &   0.04972 \\
        Sandia/adder\_dcop\_32 &  1813 &  11246 &  8 &   0.02941 &   0.01693 &                    HB/sherman2 &  1080 &  23094 &  8 &   0.08500 &   0.05379 \\
        Sandia/adder\_dcop\_33 &  1813 &  11246 &  8 &   0.02833 &   0.01710 &                       Gset/G40 &  2000 &  23532 &  8 &   1.05000 &   0.90126 \\
        Sandia/adder\_dcop\_34 &  1813 &  11246 &  8 &   0.02833 &   0.01690 &                       Gset/G39 &  2000 &  23556 &  8 &   1.03400 &   0.82907 \\
        Sandia/adder\_dcop\_35 &  1813 &  11246 &  8 &   0.02889 &   0.01693 &                       Gset/G42 &  2000 &  23558 &  8 &   1.06200 &   0.85347 \\
        Sandia/adder\_dcop\_36 &  1813 &  11246 &  8 &   0.02889 &   0.01683 &                       Gset/G41 &  2000 &  23570 &  8 &   0.99200 &   0.83307 \\
        Sandia/adder\_dcop\_37 &  1813 &  11246 &  8 &   0.03000 &   0.01693 &                     FIDAP/ex29 &  2870 &  23754 &  8 &   0.08143 &   0.04754 \\
        Sandia/adder\_dcop\_38 &  1813 &  11246 &  8 &   0.02737 &   0.01703 &                Boeing/bcsstm34 &   588 &  24270 &  8 &   0.05556 &   0.06349 \\
        Sandia/adder\_dcop\_39 &  1813 &  11246 &  8 &   0.02778 &   0.01789 &                     FIDAP/ex25 &   848 &  24369 &  8 &   0.05000 &   0.02679 \\
        Sandia/adder\_dcop\_40 &  1813 &  11246 &  8 &   0.02889 &   0.01738 &                        HB/mcfe &   765 &  24382 &  8 &   0.06500 &   0.03473 \\
        Sandia/adder\_dcop\_41 &  1813 &  11246 &  8 &   0.02941 &   0.01686 &                       Gset/G56 &  5000 &  24996 &  9 &   4.56000 &  5.42238  \\
        Sandia/adder\_dcop\_42 &  1813 &  11246 &  8 &   0.03059 &   0.01693 &                Shen/shermanACa &  3432 &  25220 &  8 &   0.14200 &   0.11558 \\
        Sandia/adder\_dcop\_43 &  1813 &  11246 &  8 &   0.03118 &   0.01696 &                     Grund/meg4 &  5860 &  25258 &  8 &   0.09667 &   0.05359 \\
        Sandia/adder\_dcop\_45 &  1813 &  11246 &  8 &   0.02941 &   0.01713 &                   HB/lns\_3937 &  3937 &  25407 &  8 &   0.18800 &   0.12218 \\
        Sandia/adder\_dcop\_46 &  1813 &  11246 &  8 &   0.02889 &   0.01700 &                    HB/lnsp3937 &  3937 &  25407 &  8 &   0.19800 &   0.12238 \\
        Sandia/adder\_dcop\_47 &  1813 &  11246 &  8 &   0.02833 &   0.01713 &                Boeing/msc01050 &  1050 &  26198 &  7 &   0.09167 &   0.01307 \\
        Sandia/adder\_dcop\_48 &  1813 &  11246 &  8 &   0.02889 &   0.01703 &                    HB/bcsstk21 &  3600 &  26600 &  7 &   0.10800 &   0.03778 \\
        Sandia/adder\_dcop\_49 &  1813 &  11246 &  8 &   0.02941 &   0.01713 &                      Bai/qc324 &   324 &  26730 & 4d &   0.02125 &   0.02182 \\
        Sandia/adder\_dcop\_50 &  1813 &  11246 &  8 &   0.02889 &   0.01670 &                      FIDAP/ex2 &   441 &  26839 &  8 &   0.02889 &   0.01900 \\
        Sandia/adder\_dcop\_51 &  1813 &  11246 &  8 &   0.02889 &   0.01670 &                       Gset/G62 &  7000 &  28000 &  8 &   0.24600 &   0.15178 \\
        Sandia/adder\_dcop\_52 &  1813 &  11246 &  8 &   0.02833 &   0.01673 &                      Hohn/fd12 &  7500 &  28462 &  8 &   0.21600 &   0.14018 \\
        Sandia/adder\_dcop\_53 &  1813 &  11246 &  8 &   0.02889 &   0.01693 &                  Grund/bayer03 &  6747 &  29195 &  8 &   0.12600 &   0.07170 \\
        Sandia/adder\_dcop\_54 &  1813 &  11246 &  8 &   0.02833 &   0.01683 &                    Bai/rdb5000 &  5000 &  29600 &  8 &   0.18400 &   0.11418 \\
        Sandia/adder\_dcop\_55 &  1813 &  11246 &  8 &   0.02889 &   0.01693 &               DRIVCAV/cavity06 &  1182 &  29675 &  8 &   0.05667 &   0.03111 \\
        Sandia/adder\_dcop\_56 &  1813 &  11246 &  8 &   0.02889 &   0.01686 &               DRIVCAV/cavity08 &  1182 &  29675 &  8 &   0.05778 &   0.02982 \\
        Sandia/adder\_dcop\_57 &  1813 &  11246 &  8 &   0.02941 &   0.01673 &                 Lucifora/cell1 &  7055 &  30082 &  8 &   0.20000 &   0.11298 \\
        Sandia/adder\_dcop\_58 &  1813 &  11246 &  8 &   0.02889 &   0.01686 &                 Lucifora/cell2 &  7055 &  30082 &  8 &   0.20000 &   0.11338 \\
        Sandia/adder\_dcop\_59 &  1813 &  11246 &  8 &   0.02833 &   0.01686 &                    HB/bcsstk26 &  1922 &  30336 &  7 &   0.13600 &   0.01638 \\
        Sandia/adder\_dcop\_60 &  1813 &  11246 &  8 &   0.03000 &   0.01690 &                      FIDAP/ex4 &  1601 &  31849 &  8 &   0.09333 &   0.05009 \\
        Sandia/adder\_dcop\_61 &  1813 &  11246 &  8 &   0.02889 &   0.01696 &                       Gset/G65 &  8000 &  32000 &  8 &   0.28600 &   0.18157 \\
        Sandia/adder\_dcop\_62 &  1813 &  11246 &  8 &   0.02889 &   0.01676 &                    HB/plat1919 &  1919 &  32399 &  7 &   0.12800 &   0.02359 \\
        Sandia/adder\_dcop\_63 &  1813 &  11246 &  8 &   0.02889 &   0.01686 &               DRIVCAV/cavity05 &  1182 &  32632 &  8 &   0.06500 &   0.03433 \\
        Sandia/adder\_dcop\_64 &  1813 &  11246 &  8 &   0.02889 &   0.01686 &                  Rajat/rajat03 &  7602 &  32653 &  8 &   0.16800 &   0.09932 \\
        Sandia/adder\_dcop\_65 &  1813 &  11246 &  8 &   0.02889 &   0.01683 &               DRIVCAV/cavity07 &  1182 &  32702 &  8 &   0.05778 &   0.03206 \\
        Sandia/adder\_dcop\_66 &  1813 &  11246 &  8 &   0.02889 &   0.01696 &               DRIVCAV/cavity09 &  1182 &  32702 &  8 &   0.06111 &   0.03206 \\
        Sandia/adder\_dcop\_67 &  1813 &  11246 &  8 &   0.02833 &   0.01670 &              Grund/poli\_large & 15575 &  33033 &  8 &   0.04333 &   0.02625 \\
        Sandia/adder\_dcop\_68 &  1813 &  11246 &  8 &   0.02889 &   0.01686 &                     HB/gemat12 &  4929 &  33044 &  8 &   0.09167 &   0.05109 \\
        Sandia/adder\_dcop\_69 &  1813 &  11246 &  8 &   0.02941 &   0.01676 &                     HB/gemat11 &  4929 &  33108 &  8 &   0.09333 &   0.05099 \\
                    HB/watt\_1 &  1856 &  11360 &  8 &   0.06000 &   0.03649 &          Hollinger/jan99jac020 &  6774 &  33744 &  8 &   0.24200 &   0.15978 \\
                    HB/watt\_2 &  1856 &  11550 &  8 &   0.07000 &   0.04015 &        Hollinger/jan99jac020sc &  6774 &  33744 &  8 &   0.24800 &   0.16877 \\
                 Grund/bayer09 &  3083 &  11767 &  8 &   0.03714 &   0.01969 &                    HB/bcsstk11 &  1473 &  34241 &  7 &   0.11000 &   0.01710 \\
                   Bai/rdb2048 &  2048 &  12032 &  8 &   0.05889 &   0.03486 &                    HB/bcsstk12 &  1473 &  34241 &  7 &   0.11200 &   0.01716 \\
                 Rajat/rajat12 &  1879 &  12818 &  8 &   0.03125 &   0.01726 &                       Gset/G61 &  7000 &  34296 &  9 &  11.28600 &  13.77691 \\
                   HB/bcsstk08 &  1074 &  12960 &  7 &   0.05556 &   0.01723 &                Boeing/msc00726 &   726 &  34518 &  7 &   0.19200 &   0.05269 \\
                  MathWorks/Pd &  8081 &  13036 &  8 &   0.01786 &   0.00994 &              Bomhof/circuit\_1 &  2624 &  35823 &  8 &   0.11200 &   0.05309 \\
                    Hamm/add20 &  2395 &  13151 &  8 &   0.04500 &   0.02485 &                       Gset/G66 &  9000 &  36000 &  8 &   0.33000 &   0.21097 \\
                 Zitney/radfr1 &  1048 &  13299 &  8 &   0.02684 &   0.01375 &                   Mallya/lhr02 &  2954 &  36875 &  8 &   0.11800 &   0.06399 \\
                  HB/orsreg\_1 &  2205 &  14133 &  8 &   0.10000 &   0.05932 &           Oberwolfach/t2dal\_a &  4257 &  37465 &  8 &   0.13800 &   0.09099 \\
       Sandia/adder\_trans\_01 &  1814 &  14579 &  8 &   0.03643 &   0.02104 &                     FIDAP/ex27 &   974 &  37652 &  8 &   0.07143 &   0.03814 \\
       Sandia/adder\_trans\_02 &  1814 &  14579 &  8 &   0.03400 &   0.02000 &                       Gset/G10 &   800 &  38352 &  8 &   0.53800 &   0.42993 \\
                   Bai/pde2961 &  2961 &  14585 &  8 &   0.07000 &   0.03807 &                        Gset/G6 &   800 &  38352 &  8 &   0.53600 &   0.42454 \\
                   HB/bcsstm25 & 15439 &  15439 & 4c &   0.00075 &   0.00248 &                        Gset/G7 &   800 &  38352 &  8 &   0.54600 &   0.44433 \\
               Boeing/bcsstm37 & 25503 &  15525 &  9 &   0.06875 &   0.02833 &                        Gset/G8 &   800 &  38352 &  8 &   0.60000 &   0.42714 \\
\hline
\end{tabular}
}
\caption{Benchmark of left-division operator on {\Matlab} R14sp2 against {\Octave} 2.9.5, on a Pentium 4M 1.6GHz machine with 1GB of memory. {\dag} The solver used for the problem, as given in section~\ref{sec:linalg}\label{tab:bench2a}}
\end{center}
\end{table*}

\begin{table*}
\begin{center}
{\tiny
\begin{tabular}{|c|c|c|c||c|c||c|c|c|c||c|c|}\hline
                 Matrix        & Order &   NNZ  & S$^\dag$ & \multicolumn{2}{|c|}{Execution Time} &
                 Matrix        & Order &   NNZ  & S$^\dag$ & \multicolumn{2}{|c|}{Execution Time} \\
                               &       &        &    &  \multicolumn{2}{|c|}{for Operator (sec)} &
                               &       &        &    &  \multicolumn{2}{|c|}{for Operator (sec)} \\ \cline{5-6} \cline{11-12}
                               &       &        &    & Matlab    &   Octave &
                               &       &        &    & Matlab    &   Octave \\ \hline
                       Gset/G9 &   800 &  38352 &  8 &   0.61400 &   0.42054 &                  Zitney/rdist1 &  4134 &  94408 &  8 &   0.16600 &   0.08565 \\
               Boeing/nasa1824 &  1824 &  39208 &  8 &   0.21000 &   0.06166 &                   Averous/epb1 & 14734 &  95053 &  8 &   0.56600 &   0.34515 \\
                 Nasa/nasa1824 &  1824 &  39208 &  7 &   0.16000 &   0.02590 &            GHS\_indef/linverse & 11999 &  95977 & 4d &   0.01378 &   0.01686 \\
                      Gset/G27 &  2000 &  39980 &  8 &   2.88000 &   2.85177 &            IBM\_Austin/coupled & 11341 &  97193 &  8 &   0.44000 &   0.23836 \\
                      Gset/G28 &  2000 &  39980 &  8 &   3.33200 &   3.14972 &                Langemyr/comsol &  1500 &  97645 &  8 &   0.15200 &   0.08449 \\
                      Gset/G29 &  2000 &  39980 &  8 &   2.76000 &   3.07973 &                Boeing/msc04515 &  4515 &  97707 &  7 &   0.53200 &   0.07527 \\
                      Gset/G30 &  2000 &  39980 &  8 &   3.07600 &   3.08493 &                     FIDAP/ex15 &  6867 &  98671 &  7 &   0.32800 &   0.07899 \\
                      Gset/G31 &  2000 &  39980 &  8 &   3.58000 &   3.06333 &                   Hamm/memplus & 17758 &  99147 &  8 &   0.74200 &   0.38054 \\
                      Gset/G67 & 10000 &  40000 &  8 &   0.38200 &   0.25376 &                      FIDAP/ex9 &  3363 &  99471 &  7 &   0.21800 &   0.04690 \\
                    HB/mbeause &   496 &  41063 &  9 &   0.20000 &   0.20537 &                  Nasa/nasa4704 &  4704 & 104756 &  7 &   0.65000 &   0.10938 \\
             vanHeukelum/cage9 &  3534 &  41594 &  8 &   0.89800 &   0.69909 &                Boeing/crystm01 &  4875 & 105339 &  7 &   0.62800 &   0.11798 \\
                    Bai/dw4096 &  8192 &  41746 &  8 &   0.34400 &   0.89946 &          Hollinger/mark3jac040 & 18289 & 106803 &  8 &  14.30200 &   7.98099 \\
               TOKAMAK/utm3060 &  3060 &  42211 &  8 &   0.17000 &   0.13018 &        Hollinger/mark3jac040sc & 18289 & 106803 &  8 &  68.61400 &   8.17816 \\
            Hollinger/g7jac020 &  5850 &  42568 &  8 &   0.65800 &   0.55212 &             Hollinger/g7jac040 & 11790 & 107383 &  8 &  22.41200 &   3.04114 \\
          Hollinger/g7jac020sc &  5850 &  42568 &  8 &   0.67800 &   0.56011 &           Hollinger/g7jac040sc & 11790 & 107383 &  8 &  23.25000 &   3.05234 \\
               Alemdar/Alemdar &  6245 &  42581 &  8 &   0.31400 &   0.23416 &          Hollinger/jan99jac060 & 20614 & 111903 &  8 &   6.11600 &   1.06424 \\
                    FIDAP/ex23 &  1409 &  42760 &  8 &   0.08500 &   0.04745 &        Hollinger/jan99jac060sc & 20614 & 111903 &  8 &   6.38000 &   1.09163 \\
                     Hohn/fd15 & 11532 &  44206 &  8 &   0.39200 &   0.26476 &                    HB/bcsstk15 &  3948 & 117816 &  7 &   7.43600 &   0.27576 \\
               Boeing/msc01440 &  1440 &  44998 &  7 &   0.17600 &   0.03992 &                  Pothen/bodyy4 & 17546 & 121550 &  7 &   3.64800 &   0.22437 \\
                   HB/bcsstk23 &  3134 &  45178 &  7 &   1.09800 &   0.22817 &                  Okunbor/aft01 &  8205 & 125567 &  9 &   1.91600 &   0.31035 \\
                     FIDAP/ex7 &  1633 &  46626 &  8 &   0.15600 &   0.08449 &                  Okunbor/aft02 &  8184 & 127762 &  8 &   3.32000 &   0.54072 \\
               Boeing/bcsstm39 & 46772 &  46772 & 4c &   0.00269 &   0.00723 &            GHS\_indef/aug3dcqp & 35543 & 128115 &  8 &  12.09600 &   7.07892 \\
                    FIDAP/ex24 &  2283 &  47901 &  8 &   0.11800 &   0.07086 &                  Pothen/bodyy5 & 18589 & 128853 &  7 &   0.78400 &   0.24776 \\
             Bomhof/circuit\_3 & 12127 &  48137 &  8 &   0.15200 &   0.08749 &                 Simon/raefsky6 &  3402 & 130371 &  8 &   0.01667 &   0.03353 \\
                 Rajat/rajat13 &  7598 &  48762 &  8 &   0.11400 &   0.06874 &             Schenk\_ISEI/igbt3 & 10938 & 130500 &  8 &   0.66000 &   0.44173 \\
                     FIDAP/ex6 &  1651 &  49062 &  9 &   0.15400 &   0.08899 &               DRIVCAV/cavity17 &  4562 & 131735 &  8 &   0.32400 &   0.18137 \\
              GHS\_indef/tuma2 & 12992 &  49365 &  8 &   0.48000 &   0.33155 &               DRIVCAV/cavity19 &  4562 & 131735 &  8 &   0.32400 &   0.18177 \\
                    HB/mbeacxc &   496 &  49920 &  9 &   0.24400 &   0.29795 &               DRIVCAV/cavity21 &  4562 & 131735 &  8 &   0.32600 &   0.18097 \\
                    HB/mbeaflw &   496 &  49920 &  9 &   0.26000 &   0.29715 &               DRIVCAV/cavity23 &  4562 & 131735 &  8 &   0.32600 &   0.18137 \\
                     FIDAP/ex3 &  1821 &  52685 &  7 &   0.11200 &   0.02291 &               DRIVCAV/cavity25 &  4562 & 131735 &  8 &   0.32600 &   0.18157 \\
         Hollinger/mark3jac020 &  9129 &  52883 &  8 &   1.68600 &   1.53417 &                  Pothen/bodyy6 & 19366 & 134208 &  7 &   0.78800 &   0.26776 \\
       Hollinger/mark3jac020sc &  9129 &  52883 &  8 &   1.73200 &   1.57276 &               DRIVCAV/cavity16 &  4562 & 137887 &  8 &   0.32400 &   0.17857 \\
                    FIDAP/ex36 &  3079 &  53099 &  8 &   0.13800 &   0.07727 &               DRIVCAV/cavity18 &  4562 & 138040 &  8 &   0.33000 &   0.18337 \\
               Shen/shermanACd &  6136 &  53329 &  8 &   0.40000 &   0.21697 &               DRIVCAV/cavity20 &  4562 & 138040 &  8 &   0.33000 &   0.18337 \\
            GHS\_indef/ncvxqp9 & 16554 &  54040 &  8 &   0.53400 &   0.35535 &               DRIVCAV/cavity22 &  4562 & 138040 &  8 &   0.33000 &   0.18337 \\
                    FIDAP/ex10 &  2410 &  54840 &  7 &   0.11200 &   0.01973 &               DRIVCAV/cavity24 &  4562 & 138040 &  8 &   0.33000 &   0.18377 \\
                   HB/bcsstk27 &  1224 &  56126 &  7 &   0.14200 &   0.01996 &               DRIVCAV/cavity26 &  4562 & 138040 &  8 &   0.32800 &   0.18337 \\
                   HB/bcsstm27 &  1224 &  56126 &  8 &   0.19400 &   0.08242 &            GHS\_indef/stokes64 & 12546 & 140034 &  9 &   2.61000 &   1.72634 \\
                 Zitney/rdist2 &  3198 &  56834 &  8 &   0.10000 &   0.04981 &           GHS\_indef/stokes64s & 12546 & 140034 &  8 &   1.25600 &   0.97105 \\
                  FIDAP/ex10hs &  2548 &  57308 &  7 &   0.14400 &   0.02171 &                    Cote/mplate &  5962 & 142190 &  8 &  40.58600 &  42.15619 \\
                    Grund/meg1 &  2904 &  58142 &  8 &   0.09333 &   0.04549 &                Shen/shermanACb & 18510 & 145149 &  8 &   0.78200 &   0.53212 \\
                      Gset/G59 &  5000 &  59140 &  8 &  11.19600 &  10.23964 &           Hollinger/g7jac050sc & 14760 & 145157 &  8 &   6.51400 &   6.19726 \\
             GHS\_indef/sit100 & 10262 &  61046 &  8 &   1.49000 &   1.36959 &                    HB/bcsstk18 & 11948 & 149090 &  7 &   2.42600 &   0.32115 \\
                Zitney/rdist3a &  2398 &  61896 &  8 &   0.10000 &   0.05229 &             GHS\_indef/bloweya & 30004 & 150009 &  8 &   1.96200 &   0.85407 \\
                 Grund/bayer02 & 13935 &  63307 &  8 &   0.31200 &   0.18137 &             vanHeukelum/cage10 & 11397 & 150645 &  8 &  33.07600 &  29.87406 \\
                     Hohn/fd18 & 16428 &  63406 &  8 &   0.61800 &   0.45653 &          Hollinger/jan99jac080 & 27534 & 151063 &  8 &   1.85400 &   1.49017 \\
                   HB/bcsstk14 &  1806 &  63454 &  7 &   0.24600 &   0.03861 &        Hollinger/jan99jac080sc & 27534 & 151063 &  8 &   2.07400 &   1.63735 \\
           LiuWenzhuo/powersim & 15838 &  64424 &  8 &   0.19200 &   0.12058 &                   Mallya/lhr07 &  7337 & 154660 &  8 &   0.52600 &   0.28616 \\
                    FIDAP/ex14 &  3251 &  65875 &  8 &   0.31600 &   0.24336 &                  Mallya/lhr07c &  7337 & 156508 &  8 &   0.51600 &   0.27816 \\
            Brunetiere/thermal &  3456 &  66528 &  8 &   0.17800 &   0.10398 &                    HB/bcsstk24 &  3562 & 159910 &  7 &   0.97600 &   0.10058 \\
                    FIDAP/ex37 &  3565 &  67591 &  8 &   0.13800 &   0.08127 &          Hollinger/mark3jac060 & 27449 & 160723 &  8 &  17.15400 &  16.08276 \\
                    FIDAP/ex20 &  2203 &  67830 &  8 &   0.15800 &   0.16557 &        Hollinger/mark3jac060sc & 27449 & 160723 &  8 &  18.11400 &  16.52869 \\
               GHS\_indef/dtoc & 24993 &  69972 &  9 &   0.09500 &   5.36218 &                     Zhao/Zhao1 & 33861 & 166453 &  8 &   6.74400 &   5.26940 \\
              GHS\_indef/aug3d & 24300 &  69984 &  9 &   0.09667 &  64.24463 &                     Zhao/Zhao2 & 33861 & 166453 &  8 &   9.35200 & 159.26146 \\
               Gaertner/nopoly & 10774 &  70842 &  8 &   0.44800 &   0.16218 &                 Simon/raefsky5 &  6316 & 167178 &  8 &   0.18400 &   0.04908 \\
                 Grund/bayer10 & 13436 &  71594 &  8 &   0.34800 &   0.20257 &             GHS\_indef/bratu3d & 27792 & 173796 &  8 &  36.78000 &  45.28932 \\
              DRIVCAV/cavity11 &  2597 &  71601 &  8 &   0.16800 &   0.09015 &                  Nasa/nasa2910 &  2910 & 174296 &  7 &   0.62800 &   0.08749 \\
              DRIVCAV/cavity13 &  2597 &  71601 &  8 &   0.16800 &   0.08982 &                   Averous/epb2 & 25228 & 175027 &  8 &   1.64800 &   1.02364 \\
              DRIVCAV/cavity15 &  2597 &  71601 &  8 &   0.16600 &   0.08999 &           Oberwolfach/t2dah\_a & 11445 & 176117 &  8 &   0.62800 &   0.39514 \\
                    FIDAP/ex18 &  5773 &  71701 &  8 &   0.22200 &   0.13758 &           Oberwolfach/t2dah\_e & 11445 & 176117 &  7 &   0.80800 &   0.18037 \\
                 Nasa/nasa2146 &  2146 &  72250 &  7 &   0.23000 &   0.04572 &                     Wang/wang3 & 26064 & 177168 &  8 &  15.15800 &  11.87799 \\
              Cannizzo/sts4098 &  4098 &  72356 &  7 &   0.39800 &   0.06010 &                     Wang/wang4 & 26068 & 177196 &  8 &  14.13600 &  11.04592 \\
         Hollinger/jan99jac040 & 13694 &  72734 &  8 &   0.71600 &   0.57891 &            GHS\_indef/brainpc2 & 27607 & 179395 &  8 &   2.68000 &   1.11383 \\
       Hollinger/jan99jac040sc & 13694 &  72734 &  8 &   0.77800 &   0.60211 &             Hollinger/g7jac060 & 17730 & 183325 &  8 &  10.86600 &   9.09742 \\
            GHS\_indef/ncvxqp1 & 12111 &  73963 &  8 &  17.04400 &  18.67276 &           Hollinger/g7jac060sc & 17730 & 183325 &  8 &   9.77600 &   8.89745 \\
                    FIDAP/ex26 &  2163 &  74464 &  8 &   0.26400 &   0.14978 &          Hollinger/jan99jac100 & 34454 & 190224 &  8 &   2.93600 &   2.18767 \\
                    FIDAP/ex13 &  2568 &  75628 &  7 &   0.15600 &   0.03250 &        Hollinger/jan99jac100sc & 34454 & 190224 &  8 &   2.97400 &   2.25206 \\
              DRIVCAV/cavity10 &  2597 &  76171 &  8 &   0.17200 &   0.09032 &          Sandia/mult\_dcop\_03 & 25187 & 193216 &  8 &   0.90600 &   0.49852 \\
              DRIVCAV/cavity12 &  2597 &  76258 &  8 &   0.17200 &   0.09165 &          Sandia/mult\_dcop\_01 & 25187 & 193276 &  8 &   1.15400 &   0.64150 \\
              DRIVCAV/cavity14 &  2597 &  76258 &  8 &   0.17200 &   0.09149 &          Sandia/mult\_dcop\_02 & 25187 & 193276 &  8 &   0.93800 &   0.50692 \\
              GHS\_indef/aug2d & 29008 &  76832 &  9 &   0.11400 &   6.86396 &            GHS\_psdef/obstclae & 40000 & 197608 &  7 &   1.42800 &   0.49213 \\
                    FIDAP/ex28 &  2603 &  77031 &  8 &   0.16800 &   0.09832 &            GHS\_psdef/torsion1 & 40000 & 197608 &  7 &   1.45800 &   0.49013 \\
                    FIDAP/ex12 &  3973 &  79077 &  9 &   0.34200 &           &            GHS\_psdef/jnlbrng1 & 40000 & 199200 &  7 &   1.57600 &   0.48233 \\
                 Gaertner/pesa & 11738 &  79566 &  8 &   0.35400 &   0.20497 &            GHS\_psdef/minsurfo & 40806 & 203622 &  7 &   1.61200 &   0.50152 \\
             GHS\_indef/aug2dc & 30200 &  80000 &  9 &   0.11800 &   7.55005 &            GHS\_indef/mario001 & 38434 & 204912 &  8 &   1.82400 &   1.20382 \\
                  Mallya/lhr04 &  4101 &  81057 &  8 &   0.26400 &   0.13998 & Schenk\_IBMSDS/2D\_27628\_bjtcai & 27628 & 206670 &  8 &   1.73400 &   1.25181 \\
                 Mallya/lhr04c &  4101 &  82682 &  8 &   0.27000 &   0.14638 &               Brethour/coater2 &  9540 & 207308 &  9 &   4.70200 &  13.37617 \\
                      Gset/G64 &  7000 &  82918 &  8 &  26.86200 &  27.66939 &          Hollinger/mark3jac080 & 36609 & 214643 &  8 &  35.83200 &  33.37233 \\
               TOKAMAK/utm5940 &  5940 &  83842 &  8 &   0.42400 &   0.33275 &        Hollinger/mark3jac080sc & 36609 & 214643 &  8 &  34.32800 &  32.84181 \\
                   HB/bcsstk13 &  2003 &  83883 &  7 &   0.78600 &   0.12238 &                    HB/bcsstk28 &  4410 & 219024 &  7 &   1.57000 &   0.12458 \\
                  Garon/garon1 &  3175 &  84723 &  8 &   0.26400 &   0.14938 &                ATandT/onetone2 & 36057 & 222596 &  8 &   1.05600 &   0.69449 \\
                    Norris/fv1 &  9604 &  85264 &  7 &   0.38000 &   0.11138 &                     FIDAP/ex35 & 19716 & 227872 &  8 &   0.75400 &   0.48453 \\
                 Grund/bayer04 & 20545 &  85537 &  9 &   0.64800 &   0.37454 &                   Mallya/lhr10 & 10672 & 228395 &  8 &   0.78800 &   0.44173 \\
                    Norris/fv2 &  9801 &  87025 &  7 &   0.51000 &   0.11518 &          Hollinger/jan99jac120 & 41374 & 229385 &  8 &   3.56400 &   3.27050 \\
                    Norris/fv3 &  9801 &  87025 &  7 &   0.52400 &   0.11498 &        Hollinger/jan99jac120sc & 41374 & 229385 &  8 &   3.80200 &   3.36849 \\
              GHS\_indef/tuma1 & 22967 &  87760 &  8 &   1.15000 &   0.93586 &            GHS\_indef/spmsrtls & 29995 & 229947 & 4d &   0.04167 &   0.04316 \\
                   HB/orani678 &  2529 &  90158 &  8 &   0.15400 &   0.08242 &                   Mallya/lhr11 & 10964 & 231806 &  9 &   1.31800 &   1.41498 \\
                     FIDAP/ex8 &  3096 &  90841 &  8 &   0.30600 &   0.17917 &                  Mallya/lhr10c & 10672 & 232633 &  8 &   0.71400 &   0.43193 \\
                    FIDAP/ex31 &  3909 &  91223 &  8 &   0.32400 &   0.17777 &                  Mallya/lhr11c & 10964 & 233741 &  8 &   0.80600 &   0.46233 \\
                  Gaertner/big & 13209 &  91465 &  8 &   0.44400 &   0.28516 &             Schenk\_ISEI/nmos3 & 18588 & 237130 &  8 &   1.76200 &   1.25961 \\
\hline
\end{tabular}
}
\caption{Benchmark of left-division operator on {\Matlab} R14sp2 against {\Octave} 2.9.5, on a Pentium 4M 1.6GHz machine with 1GB of memory. {\dag} The solver used for the problem, as given in section~\ref{sec:linalg}.\label{tab:bench2b}}
\end{center}
\end{table*}

\begin{table*}
\begin{center}
{\tiny
\begin{tabular}{|c|c|c|c||c|c||c|c|c|c||c|c|}\hline
                 Matrix        & Order &   NNZ  & S$^\dag$ & \multicolumn{2}{|c|}{Execution Time} &
                 Matrix        & Order &   NNZ  & S$^\dag$ & \multicolumn{2}{|c|}{Execution Time} \\
                               &       &        &    &  \multicolumn{2}{|c|}{for Operator (sec)} &
                               &       &        &    &  \multicolumn{2}{|c|}{for Operator (sec)} \\ \cline{5-6} \cline{11-12}
                               &       &        &    & Matlab    &   Octave &
                               &       &        &    & Matlab    &   Octave \\ \hline
                   HB/bcsstk25 & 15439 & 252241 &  7 &   4.33800 &   0.64130 &                Nemeth/nemeth12 &  9506 & 446818 & 4d &   1.26000 &   0.12823 \\
           GHS\_indef/a5esindl & 60008 & 255004 &  8 &   6.65000 &   2.84177 &                     FIDAP/ex40 &  7740 & 456188 &  8 &  10.35600 &   1.55876 \\
                    FIDAP/ex19 & 12005 & 259577 &  8 &   0.54000 &   0.31515 &                    Bai/af23560 & 23560 & 460598 &  8 &   5.89000 &   4.56431 \\
            Hollinger/g7jac080 & 23670 & 259648 &  8 &  18.99200 &  18.32181 &                     Bai/qc2534 &  2534 & 463360 & 4d &   0.91600 &   0.93286 \\
          Hollinger/g7jac080sc & 23670 & 259648 &  8 &  18.56000 &  18.23663 &                   Averous/epb3 & 84617 & 463625 &  8 &   3.63400 &   2.49002 \\
         Hollinger/mark3jac100 & 45769 & 268563 &  8 &  36.84000 &  35.18085 &           GHS\_psdef/wathen100 & 30401 & 471601 &  7 &   3.16800 &   0.58271 \\
       Hollinger/mark3jac100sc & 45769 & 268563 &  8 &  35.15400 &  33.50571 &                Nemeth/nemeth13 &  9506 & 474472 & 4d &   0.13400 &   0.12898 \\
                 Grund/bayer01 & 57735 & 275094 &  8 &   1.41000 &   0.85467 &                GHS\_indef/c-59 & 41282 & 480536 &  8 &  14.85000 &  20.50428 \\
                   Hohn/sinc12 &  7500 & 283992 &  8 &  22.16000 &  21.10159 & Schenk\_IBMSDS/2D\_54019\_highK & 54019 & 486129 &  8 &   5.37800 &   3.79222 \\
Schenk\_IBMSDS/3D\_28984\_Tetra & 28984 & 285092 &  9 &  69.97200 & 231.32383 &             Hollinger/g7jac140 & 41490 & 488633 &  8 &  46.23200 &  45.04975 \\
                   HB/bcsstk16 &  4884 & 290378 &  7 &   3.80600 &   0.36494 &           Hollinger/g7jac140sc & 41490 & 488633 &  8 &  42.14800 &  42.86328 \\
                Simon/raefsky1 &  3242 & 293409 &  8 &   2.09000 &   1.61095 &                   Norris/lung2 & 109460 & 492564 &  8 &   1.65200 &   1.12763 \\
                Simon/raefsky2 &  3242 & 293551 &  8 &   1.93600 &   1.55796 &                Nemeth/nemeth14 &  9506 & 496144 & 4d &   0.10600 &   0.14248 \\
           GHS\_indef/dixmaanl & 60000 & 299998 &  8 &  43.68600 &   2.24486 &            Oberwolfach/t3dl\_a & 20360 & 509866 &  8 &  38.20000 &  70.31051 \\
                 Cote/vibrobox & 12328 & 301700 &  9 &  65.21000 & 628.86960 &            GHS\_psdef/gridgena & 48962 & 512084 &  7 &   5.88000 &   1.10463 \\
            FEMLAB/waveguide3D & 21036 & 303468 &  8 &   7.07400 &   6.10327 &                  Hamm/hcircuit & 105676 & 513072 &  8 &   2.47333 &   1.55656 \\
                  Mallya/lhr14 & 14270 & 305750 &  9 &   1.82400 &   2.11008 &             Schenk\_IBMNA/c-67 & 57975 & 530229 &  8 &   4.88333 &   2.67899 \\
             Bomhof/circuit\_4 & 80209 & 307604 &  8 &   6.63400 &   3.35369 &   Schenk\_IBMSDS/3D\_51448\_3D & 51448 & 537038 &  8 &  35.05333 &  30.47137 \\
                 Mallya/lhr14c & 14270 & 307858 &  8 &   1.03000 &   0.62830 &  Schenk\_IBMSDS/ibm\_matrix\_2 & 51448 & 537038 &  8 &  35.02667 &  30.68494 \\
               Boeing/crystk01 &  4875 & 315891 &  8 &   4.25400 &   0.92006 &                Nemeth/nemeth15 &  9506 & 539802 & 4d &   0.14250 &   0.17264 \\
               Boeing/bcsstm36 & 23052 & 320606 &  9 &   1.76000 &   1.40079 &                   HB/psmigr\_2 &  3140 & 540022 &  8 &  15.96000 &  15.07951 \\
         Hollinger/mark3jac120 & 54929 & 322483 &  8 &  56.95800 &  58.56690 &                   HB/psmigr\_1 &  3140 & 543160 &  8 &  13.55333 &  11.82780 \\
       Hollinger/mark3jac120sc & 54929 & 322483 &  8 &  54.82000 &  49.60506 &                   HB/psmigr\_3 &  3140 & 543160 &  8 &  13.55667 &  11.81980 \\
               Boeing/crystm02 & 13965 & 322905 &  7 &   7.49000 &   1.15942 &                    Hohn/sinc15 & 11532 & 551184 &  8 &  74.47667 &  76.96890 \\
               Goodwin/goodwin &  7320 & 324772 &  8 &   1.10400 &   1.82432 &                GHS\_indef/c-58 & 37595 & 552551 &  8 &  18.32333 &  17.36596 \\
                  Shyy/shyy161 & 76480 & 329762 &  8 &   3.04600 &   1.86832 &                  Shen/e40r0100 & 17281 & 553562 &  8 &   2.46333 &   1.90291 \\
                Graham/graham1 &  9035 & 335472 &  8 &   3.63400 &   1.21661 &             GHS\_indef/c-62ghs & 41731 & 559339 &  8 & 118.40667 & 191.48309 \\
               ATandT/onetone1 & 36057 & 335552 &  8 &   4.42400 &   4.15777 &             GHS\_indef/k1\_san & 67759 & 559774 &  9 &  62.08000 & 203.85401  \\
            Hollinger/g7jac100 & 29610 & 335972 &  8 & 129.29400 &  26.32780 &             Hollinger/g7jac160 & 47430 & 564952 &  8 &  54.45333 &  54.55504 \\
          Hollinger/g7jac100sc & 29610 & 335972 &  8 &  23.99000 &  24.64405 &           Hollinger/g7jac160sc & 47430 & 564952 &  8 &  52.34667 &  50.83061 \\
           Oberwolfach/gyro\_m & 17361 & 340431 &  7 &   1.74600 &   0.20337 &           GHS\_psdef/wathen120 & 36441 & 565761 &  7 &   3.65333 &   0.72822 \\
           GHS\_indef/a2nnsnsl & 80016 & 347222 &  8 &  11.87000 &   5.25320 &                GHS\_indef/c-68 & 64810 & 565996 &  8 &  64.63667 & 105.85824 \\
           GHS\_indef/ncvxbqp1 & 50000 & 349968 &  8 &  14.00200 &  57.09932 &                Boeing/crystm03 & 24696 & 583770 &  7 &  15.32667 &   2.95222 \\
            GHS\_psdef/cvxbqp1 & 50000 & 349968 &  7 & 288.96000 &   1.81272 &                Nemeth/nemeth16 &  9506 & 587012 & 4d &   0.15000 &   0.17631 \\
             FEMLAB/poisson3Da & 13514 & 352762 &  8 &   9.34600 &   7.84581 &                Mulvey/finan512 & 74752 & 596992 &  7 &  60.97333 &   1.66041 \\
           GHS\_indef/a0nsdsil & 80016 & 355034 &  8 &  11.69200 &   5.56075 &                GHS\_indef/c-69 & 67458 & 623914 &  8 &  20.12000 &  21.79102 \\
               Boeing/bcsstk38 &  8032 & 355460 &  7 &   4.43000 &   0.31395 &                Nemeth/nemeth17 &  9506 & 629620 & 4d &   0.15750 &   0.17564 \\
                  Garon/garon2 & 13535 & 373235 &  8 &   1.72400 &   1.19242 &            GHS\_indef/blockqp1 & 60012 & 640033 &  8 &  19.80667 &   7.89080 \\
                 Hamm/bcircuit & 68902 & 375558 &  8 &   1.68000 &   1.46478 &             Hollinger/g7jac180 & 53370 & 641290 &  8 &  69.42000 &  73.73179 \\
         Hollinger/mark3jac140 & 64089 & 376395 &  8 & 117.82000 & 117.09100 &           Hollinger/g7jac180sc & 53370 & 641290 &  8 &  64.18667 &  67.95934 \\
       Hollinger/mark3jac140sc & 64089 & 376395 &  8 & 221.90400 & 112.53149 &                GHS\_indef/c-70 & 68924 & 658986 &  8 &  21.58000 &  22.26295 \\
           Cunningham/k3plates & 11107 & 378927 &  8 &   0.93400 &   0.60891 &                  Norris/heart2 &  2339 & 680341 &  8 &   1.47667 &   1.05984 \\
                  Mallya/lhr17 & 17576 & 379761 &  9 &   2.02800 &   2.72539 &                  Norris/heart3 &  2339 & 680341 &  8 &   1.46333 &   1.05284 \\
                Sanghavi/ecl32 & 51993 & 380415 &  8 & 328.74800 & 282.02393 &                Nemeth/nemeth18 &  9506 & 695234 & 4d &   0.21333 &   0.18231 \\
                 Mallya/lhr17c & 17576 & 381975 &  8 &   1.29800 &   0.81428 &                GHS\_indef/c-72 & 84064 & 707546 &  8 &  17.22667 &  19.39872 \\
               Nemeth/nemeth02 &  9506 & 394808 &  8 &   0.54600 &   0.32155 &             Schenk\_IBMNA/c-64 & 51035 & 707985 &  8 &   7.26667 &   4.05572 \\
               Nemeth/nemeth03 &  9506 & 394808 &  8 &   0.58000 &   0.32115 &           ACUSIM/Pres\_Poisson & 14822 & 715804 &  7 &   7.04667 &   1.13183 \\
               Nemeth/nemeth04 &  9506 & 394808 &  8 &   0.54800 &   0.34095 &             Hollinger/g7jac200 & 59310 & 717620 &  8 &  81.66000 &  82.95206 \\
               Nemeth/nemeth05 &  9506 & 394808 &  8 &   0.58200 &   0.32015 &           Hollinger/g7jac200sc & 59310 & 717620 &  8 &  76.65333 &  78.69804 \\
               Nemeth/nemeth06 &  9506 & 394808 &  8 &   0.57800 &   0.33935 &                Nemeth/nemeth01 &  9506 & 725054 & 4d &   0.24000 &   0.22097 \\
               Nemeth/nemeth07 &  9506 & 394812 &  8 &   0.58200 &   0.33715 &            GHS\_indef/olesnik0 & 88263 & 744216 &  8 &  42.14000 &  42.00061 \\
               Nemeth/nemeth08 &  9506 & 394816 &  8 &   0.58000 &   0.33455 &                   Mallya/lhr34 & 35152 & 746972 &  9 &   3.12333 &   4.82847 \\
               Nemeth/nemeth09 &  9506 & 395506 &  8 &   0.57400 &   0.32995 &             GHS\_indef/copter2 & 55476 & 759952 &  8 &  37.76333 & 197.47398 \\
               Nemeth/nemeth10 &  9506 & 401448 &  8 &   0.56400 &   0.32135 &                Andrews/Andrews & 60000 & 760154 &  7 &   NC      &  60.61812 \\
               GHS\_indef/c-55 & 32780 & 403450 &  8 &  55.78800 &  72.13503 &                  Mallya/lhr34c & 35152 & 764014 &  8 &   3.39000 &   1.81506 \\
               Nemeth/nemeth11 &  9506 & 408264 & 4d &   0.56600 &   0.12723 &                Nemeth/nemeth19 &  9506 & 818302 & 4d &   0.28333 &   0.20064 \\
            Hollinger/g7jac120 & 35550 & 412306 &  8 & 188.60800 &  50.74469 &                  FEMLAB/sme3Da & 12504 & 874887 &  8 &   2.57667 &   2.17434 \\
          Hollinger/g7jac120sc & 35550 & 412306 &  8 &  46.90000 &  49.04154 &   Schenk\_IBMSDS/matrix-new\_3 & 125329 & 893984 &  8 &  65.39000 &  58.33580 \\
            GHS\_indef/ncvxqp5 & 62500 & 424966 &  8 & 537.44800 & 337.53729 &                       Kim/kim1 & 38415 & 933195 &  8 &  15.63333 &  13.86589 \\
           GHS\_indef/helm3d01 & 32226 & 428444 &  8 &  25.22600 &  60.50040 &                    Hohn/sinc18 & 16428 & 948696 &  8 & 580.02000 & 221.20670 \\
                   HB/bcsstk17 & 10974 & 428650 &  7 &   5.04200 &   0.41694 &                  Hamm/scircuit & 170998 & 958936 &  8 &   6.67000 &   6.49701 \\
               GHS\_indef/c-63 & 44234 & 434704 &  8 &   9.09400 &  10.47781 &                Boeing/crystk02 & 13965 & 968583 &  8 &  36.66333 &  12.20081 \\
           GHS\_indef/cont-201 & 80595 & 438795 &  8 &  70.84000 &  11.44966 &                Nemeth/nemeth20 &  9506 & 971870 & 4d &   0.27000 &   0.24696 \\
            Schenk\_IBMNA/c-66 & 49989 & 444853 &  8 &  51.27200 &   4.57051 &            GHS\_indef/cont-300 & 180895 & 988195 &  8 &  35.95333 &  34.57574 \\
\hline
\end{tabular}
}
\caption{Benchmark of left-division operator on {\Matlab} R14sp2 against {\Octave} 2.9.5, on a Pentium 4M 1.6GHz machine with 1GB of memory. {\dag} The solver used for the problem, as given in section~\ref{sec:linalg}.\label{tab:bench2c}}
\end{center}
\end{table*}

For the banded problems both {\Octave} and {\Matlab} perform
similarly, with only minor differences, probably due to the fact that
the same ATLAS library was used. {\Matlab} is slightly faster for
problems with very short run times, probably for similar reasons
as for small multiplications and additions.

One class of problems where the speed of {\Octave} significantly
exceeds that of {\Matlab} are the positive definite matrices that are
not solved with the LAPACK banded solvers (xPTSV or xPBTRF). This is
due in large part to the use of CHOLMOD~\cite{davis.2006} together
with the use of METIS~\cite{metis.website} for the graph partitioning.
As CHOLMOD will become the sparse Cholesky solver in future versions
of {\Matlab}~\footnote{Tim Davis has stated ``CHOLMOD will become
x=A$\backslash$b in a future release of {\Matlab} when A is symmetric and
positive definite or Hermitian, with a speedup of 5 (for small
matrices) to 40 (for big ones), depending on the matrix''} this
situation is a temporary advantage for {\Octave}. The worst case for
this is the \textit{Andrews/Andrews} matrix, where {\Matlab} did not
complete the solution due to a lack of memory. Once {\Matlab} uses
CHOLMOD, it might be expected that in this case as well similar speeds
might be expected.

The differences in the problems solved via LU decomposition using
UMFPACK are harder to explain. There are a couple of very large
discrepancies in the results, with {\Octave} winning in some cases
(cf. \textit{Hollinger/g7jac100}) and {\Matlab} in others (cf
\textit{Zhao/Zhao2}).

Both {\Octave} and {\Matlab} use recent versions of UMFPACK, with
{\Octave} using a slightly newer version to allow the use of C99
compatible complex numbers where the real and imaginary parts are
stored together. There are however no changes between the versions of
UMFPACK used that would explain any performance differences.
{\Octave} has a slight advantage when the arguments are complex, due
it is use of C99 compatible complex as it is this format that is used
internally to UMFPACK. Another possible source of differences is that
UMFPACK calls internally a column reordering routine, and {\Octave}
uses this functionality Perhaps {\Matlab} attempts to independently
guess an initial column reordering. In any case, in 11 of the cases
where UMFPACK was used the speed of {\Matlab} exceeded the speed of
{\Octave}, while in 267 of the cases the speed of {\Octave} exceeded
the speed of {\Matlab}, with the mean speed of {\Octave} being 12\%
above that of {\Matlab}.

Finally, there are significant differences between the results for
{\Octave} and {\Matlab} for singular matrices. The major difference is
that {\Matlab} uses Given's rotations whereas {\Octave} uses
Householder reflections. Given's rotations of {\Matlab} allow row
reordering to be performed to reduce the amount of work to below that
of a Householder transformation. However, the underlying code used in
{\Octave} uses Householder transformation to allow the eventual use of
multi-frontal techniques to the QR factorization, and so this option
is not available to {\Octave} currently.

Furthermore, {\Octave} uses a Dulmage-Mendelsohn
factorization of the matrix to allow the problems to be solved as a 
combination of over-determined, well-determined and under-determined
parts. The advantage of this is the potential for significantly better
performance and more stable results for over-determined
problems. However, it is possible that the Dulmage-Mehdelsohn
factorization identifies no useful structure. A case where this occurs
is the \textit{GHS\_indef/dtoc} matrix where 3 times the computation
time of a straight QR solution is needed.

The Dulmage-Mendelsohn solver can be bypassed with code like

{\small
\begin{lstlisting}
  [c,r] = qr(a,b);
  x = r\c;
\end{lstlisting}
}

It should be noted that both {\Octave} and {\Matlab} use accelerated
algorithms for the left-division operator for triangular, permuted
triangular and tridiagonal matrices, as discussed in
section~\ref{sec:linalg}, and that these cases are not treated in the
matrices from the University of Florida collection used here. These
are trivial cases, but important in that they should not be solved
with generic code.

\section{Use of Octave Sparse Matrices in Real Life Example}
\label{sec:example}

A common application for sparse matrices is in the solution of
Finite Element Models. Finite element models allow numerical
solution of partial differential equations that do not have
closed form solutions, typically because of the complex shape
of the domain.

In order to motivate this application, we consider the boundary
value Laplace equation. This system can model scalar potential
fields, such as heat or electrical potential. Given a medium
$\Omega$ with boundary $\partial\Omega$. At all points on the
$\partial\Omega$ the boundary conditions are known, and we wish
to calculate the potential in $\Omega$. Boundary conditions
may specify the potential (Dirichlet boundary condition),
its normal derivative across the boundary (Neumann boundary
condition), or a weighted sum of the potential and its
derivative (Cauchy boundary condition).

In a thermal model, we want to calculate the temperature in
$\Omega$ and know the boundary temperature (Dirichlet condition)
or heat flux (from which we can calculate the Neumann condition
by dividing by the thermal conductivity 
 at the boundary). Similarly,
in an electrical model, we want to calculate the voltage in
$\Omega$ and know the boundary voltage (Dirichlet) or current
(Neumann condition after diving by the electrical conductivity).
In an electrical model, it is common for much of the boundary
to be electrically isolated; this is a Neumann boundary condition
with the current equal to zero.

The simplest finite element models will divide $\Omega$ into 
simplexes (triangles in 2D, pyramids in 3D).  A 3D example
is shown in Figure~\ref{fig:eidorsball}, and represents a cylindrical
liquid filled tank with a small non-conductive
ball~\cite{eidors.website,eidors2006}. This is model is designed to 
reflect an application of electrical impedance tomography, where
current patterns are applied to such a tank in order to 
image the internal conductivity distribution. In order to
describe the FEM geometry, we have a matrix of vertices 
{\tt nodes} and simplices {\tt elems}.

The following example creates a simple rectangular 2D electrically
conductive medium with 10~V and 20~V imposed on opposite sides 
(Dirichlet boundary conditions). All other edges are electrically
isolated.

{\small
\begin{lstlisting}
   node_y= [1;1.2;1.5;1.8;2]*ones(1,11);
   node_x= ones(5,1)*[1,1.05,1.1,1.2, ...
             1.3,1.5,1.7,1.8,1.9,1.95,2];
   nodes= [node_x(:), node_y(:)];

   [h,w]= size(node_x);
   elems= [];
   for idx= 1:w-1
     widx= (idx-1)*h;
     elems= [elems; ...
       widx+[(1:h-1);(2:h);h+(1:h-1)]'; ...
       widx+[(2:h);h+(2:h);h+(1:h-1)]' ]; 
   endfor

   E= size(elems,1); # No. of simplices
   N= size(nodes,1); # No. of vertices
   D= size(elems,2); # dimensions+1
\end{lstlisting}
}

This creates a $N{\times}2$ matrix {\tt nodes} and a
$E\times3$ matrix {\tt elems} with values, which define
finite element triangles:

{\small
\begin{lstlisting}
  nodes(1:7,:)'
    1.00 1.00 1.00 1.00 1.00 1.05 1.05 ...
    1.00 1.20 1.50 1.80 2.00 1.00 1.20 ...

  elems(1:7,:)'
    1    2    3    4    2    3    4 ...
    2    3    4    5    7    8    9 ...
    6    7    8    9    6    7    8 ...
\end{lstlisting}
}

Using a first order FEM, we approximate the electrical
conductivity distribution in $\Omega$ as constant
on each simplex
(represented by the vector {\tt conductivity}).
Based on the finite element geometry, we first
calculate a system (or stiffness) matrix for each
simplex (represented as $3\times3$ elements on the
diagonal of the element-wise system matrix {\tt SE}.
Based on {\tt SE} and a $N{\times}DE$ connectivity
matrix {\tt C},
representing the connections between simplices and
vectices, the global connectivity matrix {\tt S} is
calculated.

{\small
\begin{lstlisting}
  # Element conductivity
  conductivity= [1*ones(1,16), ...
         2*ones(1,48), 1*ones(1,16)];

  # Connectivity matrix
  C = sparse ((1:D*E), reshape (elems', ...
         D*E, 1), 1, D*E, N);

  # Calculate system matrix
  Siidx = floor ([0:D*E-1]'/D) * D * ...
         ones(1,D) + ones(D*E,1)*(1:D) ;
  Sjidx = [1:D*E]'*ones(1,D);
  Sdata = zeros(D*E,D);
  dfact = factorial(D-1);
  for j=1:E
     a = inv([ones(D,1), ... 
         nodes(elems(j,:), :)]);
     const = conductivity(j) * 2 / ...
         dfact / abs(det(a));
     Sdata(D*(j-1)+(1:D),:) = const * ...
         a(2:D,:)' * a(2:D,:);
  endfor
  # Element-wise system matrix
  SE= sparse(Siidx,Sjidx,Sdata);
  # Global system matrix
  S= C'* SE *C;
\end{lstlisting}
}

The system matrix acts like the conductivity $S$ in Ohm's
law $SV = I$. Based on the Dirichlet and Neumann boundary
conditions, we are able to solve for the voltages at each
vertex $V$. 

{\small
\begin{lstlisting}
  # Dirichlet boundary conditions
  D_nodes=[1:5, 51:55]; 
  D_value=[10*ones(1,5), 20*ones(1,5)]; 

  V= zeros(N,1);
  V(D_nodes) = D_value;
  idx = 1:N; # vertices without Dirichlet 
             # boundary condns
  idx(D_nodes) = [];

  # Neumann boundary conditions. Note that
  # N_value must be normalized by the
  # boundary length and element conductivity
  N_nodes=[];
  N_value=[];

  Q = zeros(N,1);
  Q(N_nodes) = N_value;

  V(idx) = S(idx,idx) \ ( Q(idx) - ...
            S(idx,D_nodes) * V(D_nodes));
\end{lstlisting}
}

\begin{figure}
  \begin{minipage}[b]{1.0\linewidth}
    \centering
    \centerline{\epsfig{figure=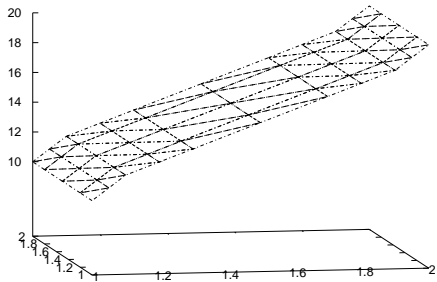,width=8.5cm}}
  \end{minipage}
  \begin{center}
  \caption{Example finite element model
           the showing triangular elements.
           The height of each vertex corresponds to the 
           solution value}
  \label{fig:fem_model}
  \end{center}
\end{figure}

Finally, in order to display the solution, we show each
solved voltage value in the z-axis for each simplex vertex
in Figure~\ref{fig:fem_model}.

{\small
\begin{lstlisting}
  elemx = elems(:,[1,2,3,1])';
  xelems = reshape (nodes(elemx, 1), 4, E);
  yelems = reshape (nodes(elemx, 2), 4, E);
  velems = reshape (V(elemx), 4, E);
  plot3 (xelems,yelems,velems,'k'); 
  print ('grid.eps');
\end{lstlisting}
}

\section{Using Sparse Matrices in Oct-Files}
\label{sec:octfiles}

An {\OctFile} is a means of writing an {\Octave} function in a
compilable language like C++, rather than as a script file. This can
result in a significant acceleration in the code.  It is not the
purpose of this section to discuss how to write an {\OctFile}, or
discuss what they are. Users wishing to find out more about
{\OctFiles} themselves are referred to the articles by Cristophe
Spiel~\cite{spiel.del_coda} and Paul Thomas~\cite{thomas.dal_segno}.
Users who are not familiar with {\OctFiles} are urged to read these
references to fully understand this section. The examples discussed
here assume that the {\OctFile} is written entirely in C++.

There are three classes of sparse objects that are of interest to the
user.

\begin{itemize}
\item SparseMatrix - double precision sparse matrix class

\item SparseComplexMatrix - Complex sparse matrix class

\item SparseBoolMatrix - boolean sparse matrix class
\end{itemize}

All of these classes inherit from the \textit{Sparse$<$T$>$} template
class, and so all have similar capabilities and usage. The
\textit{Sparse$<$T$>$} class was based on \textit{Array$<$T$>$} 
class, and so users familiar with {\Octave}'s array classes will be
comfortable with the use of the sparse classes.

The sparse classes will not be entirely described in this section, due
to their similar with the existing array classes. However, there are a
few differences due the different nature of sparse objects, and these
will be described. Firstly, although it is fundamentally possible to
have N-dimensional sparse objects, the {\Octave} sparse classes do not
allow them at this time. So all operations of the sparse classes must
be 2-dimensional.  This means that in fact \textit{SparseMatrix} is
similar to {\Octave}'s \textit{Matrix} class rather than its
\textit{NDArray} class.

\subsection{Differences between the Array and Sparse Classes}
\label{sec:octdif}

The number of elements in a sparse matrix is considered to be the number
of non-zero elements rather than the product of the dimensions. Therefore

{\small
\begin{lstlisting}
  SparseMatrix sm;
  ...
  int nel = sm.nelem ();
\end{lstlisting}
}

returns the number of non-zero elements. If the user really requires
the number of elements in the matrix, including the non-zero elements,
they should use \textit{numel} rather than \textit{nelem}. Note that
for very large matrices, where the product of the two dimensions is
larger than the representation of the an \textit{octave\_idx\_type}, then
\textit{numel} can overflow.  An example is \textit{speye(1e6)} which will
create a matrix with a million rows and columns, but only a million
non-zero elements. Therefore the number of rows by the number of
columns in this case is more than two hundred times the maximum value
that can be represented by an unsigned int on a 32-bit platform.  The
use of \textit{numel} should therefore be avoided useless it is known
it won't overflow.

Extreme care must be taken with the elem method and the \textit{()}
operator, which perform basically the same function. The reason is
that if a sparse object is non-const, then {\Octave} will assume that a
request for a zero element in a sparse matrix is in fact a request to
create this element so it can be filled. Therefore a piece of code
like

{\small
\begin{lstlisting}
  SparseMatrix sm;
  ...
  for (int j = 0; j < nc; j++)
    for (int i = 0; i < nr; i++)
      std::cerr << " (" << i << "," 
                << j << "): " << sm(i,j) 
                << std::endl;
\end{lstlisting}
}

is a great way of turning the sparse matrix into a dense one, and a
very slow way at that since it reallocates the sparse object at each
zero element in the matrix.

An easy way of preventing the above from happening is to create a
temporary constant version of the sparse matrix. Note that only the
container for the sparse matrix will be copied, while the actual
representation of the data will be shared between the two versions of
the sparse matrix. So this is not a costly operation. For example, the
above would become

{\small
\begin{lstlisting}
  SparseMatrix sm;
  ...
  const SparseMatrix tmp (sm);
  for (int j = 0; j < nc; j++)
    for (int i = 0; i < nr; i++)
      std::cerr << " (" << i << "," 
                << j << "): " << tmp(i,j) 
                << std::endl;
\end{lstlisting}
}

Finally, as the sparse types aren't just represented as a contiguous
block of memory, the \textit{fortran\_vec} method of the
\textit{Array$<$T$>$} class is not available. It is however replaced by three
separate methods \textit{ridx}, \textit{cidx} and \textit{data}, that
access the raw compressed column format that the {\Octave} sparse
matrices are stored in.  Additionally, these methods can be used in a
manner similar to \textit{elem}, to allow the matrix to be accessed or
filled. However, in that case it is up to the user to respect the
sparse matrix compressed column format discussed previous.

\subsection{Creating Spare Matrices in Oct-Files}
\label{sec:octcreate}

The user has several alternatives in how to create a sparse matrix.
They can first create the data as three vectors representing the
row and column indexes and the data, and from those create the matrix.
Or alternatively, they can create a sparse matrix with the appropriate
amount of space and then fill in the values. Both techniques have their
advantages and disadvantages.

An example of how to create a small sparse matrix with the first technique
might be seen the example

{\small
\begin{lstlisting}
  int nz = 4, nr = 3, nc = 4;
  ColumnVector ridx (nz);
  ColumnVector cidx (nz);
  ColumnVector data (nz);

  ridx(0) = 0; ridx(1) = 0; 
  ridx(2) = 1; ridx(3) = 2;
  cidx(0) = 0; cidx(1) = 1; 
  cidx(2) = 3; cidx(3) = 3;
  data(0) = 1; data(1) = 2; 
  data(2) = 3; data(3) = 4;

  SparseMatrix sm(data, ridx, cidx, nr, nc);
\end{lstlisting}
}

which creates the matrix given in section~\ref{sec:storage}. Note that
the compressed matrix format is not used at the time of the creation
of the matrix itself, however it is used internally.

As previously mentioned, the values of the sparse matrix are stored
in increasing column-major ordering. Although the data passed by the
user does not need to respect this requirement, the pre-sorting the
data significantly speeds up the creation of the sparse matrix.

The disadvantage of this technique of creating a sparse matrix is that
there is a brief time where two copies of the data exists. Therefore
for extremely memory constrained problems this might not be the right
technique to create the sparse matrix.

The alternative is to first create the sparse matrix with the desired
number of non-zero elements and then later fill those elements in. The
easiest way to do this is

{\small
\begin{lstlisting}
  int nz = 4, nr = 3, nc = 4;
  SparseMatrix sm (nr, nc, nz);
  sm(0,0) = 1; sm(0,1) = 2; 
  sm(1,3) = 3; sm(2,3) = 4;
\end{lstlisting}
}

That creates the same matrix as previously. Again, although it is not
strictly necessary, it is significantly faster if the sparse matrix is
created in this manner that the elements are added in column-major
ordering. The reason for this is that if the elements are inserted
at the end of the current list of known elements then no element
in the matrix needs to be moved to allow the new element to be
inserted. Only the column indexes need to be updated.

There are a few further points to note about this technique of creating
a sparse matrix. Firstly, it is not illegal to create a sparse matrix 
with fewer elements than are actually inserted in the matrix. Therefore

{\small
\begin{lstlisting}
  int nz = 4, nr = 3, nc = 4;
  SparseMatrix sm (nr, nc, 0);
  sm(0,0) = 1; sm(0,1) = 2; 
  sm(1,3) = 3; sm(2,3) = 4;
\end{lstlisting}
}

is perfectly legal, but will be very slow. The reason is that 
as each new element is added to the sparse matrix the space allocated
to it is increased by reallocating the memory. This is an expensive
operation, that will significantly slow this means of creating a sparse
matrix. Furthermore, it is not illegal to create a sparse matrix with 
too much storage, so having \textit{nz} above equaling 6 is also legal.
The disadvantage is that the matrix occupies more memory than strictly
needed.

It is not always easy to know the number of non-zero elements prior
to filling a matrix. For this reason the additional storage for the
sparse matrix can be removed after its creation with the
\textit{maybe\_compress} function. Furthermore, 
\textit{maybe\_compress} can deallocate the unused storage, but it 
can equally remove zero elements from the matrix.  The removal of zero
elements from the matrix is controlled by setting the argument of the
\textit{maybe\_compress} function to be 'true'. However, the cost of
removing the zeros is high because it implies resorting the
elements. Therefore, if possible it is better is the user doesn't add
the zeros in the first place. An example of the use of
\textit{maybe\_compress} is

{\small
\begin{lstlisting}
  int nz = 6, nr = 3, nc = 4;
  SparseMatrix sm1 (nr, nc, nz);
  sm1(0,0) = 1; sm1(0,1) = 2; 
  sm1(1,3) = 3; sm1(2,3) = 4;
  // No zero elements were added
  sm1.maybe_compress ();

  SparseMatrix sm2 (nr, nc, nz);
  sm2(0,0) = 1; sm2(0,1) = 2; 
  sm2(0,2) = 0; sm2(1,2) = 0; 
  sm2(1,3) = 3; sm2(2,3) = 4;
  // Zero elements were added
  sm2.maybe_compress (true);
\end{lstlisting}
}

The \textit{maybe\_compress} function should be avoided if
possible, as it will slow the creation of the matrices.

A third means of creating a sparse matrix is to work directly with
the data in compressed row format. An example of this technique might
be

{\small
\begin{lstlisting}
  octave_value arg;
  ...

  // Assume we know the max no nz 
  int nz = 6, nr = 3, nc = 4;
  SparseMatrix sm (nr, nc, nz);
  Matrix m = arg.matrix_value ();

  int ii = 0;
  sm.cidx (0) = 0;
  for (int j = 1; j < nc; j++)
    {
      for (int i = 0; i < nr; i++)
        {
          double tmp = foo (m(i,j));
          if (tmp != 0.)
            {
              sm.data(ii) = tmp;
              sm.ridx(ii) = i;
              ii++;
            }
        }
      sm.cidx(j+1) = ii;
   }
  // Don't know a-priori the final no of nz.
  sm.maybe_compress (); 
\end{lstlisting}
}

which is probably the most efficient means of creating the sparse matrix.

Finally, it might sometimes arise that the amount of storage initially
created is insufficient to completely store the sparse matrix. Therefore,
the method \textit{change\_capacity} exists to reallocate the sparse 
memory. The above example would then be modified as 

{\small
\begin{lstlisting}
  octave_value arg;
  ... 

  // Assume we know the max no nz 
  int nz = 6, nr = 3, nc = 4;
  SparseMatrix sm (nr, nc, nz);
  Matrix m = arg.matrix_value ();

  int ii = 0;
  sm.cidx (0) = 0;
  for (int j = 1; j < nc; j++)
    {
      for (int i = 0; i < nr; i++)
        {
          double tmp = foo (m(i,j));
          if (tmp != 0.)
            {
              if (ii == nz)
                {
                  // Add 2 more elements
                  nz += 2;
                  sm.change_capacity (nz);
                }
              sm.data(ii) = tmp;
              sm.ridx(ii) = i;
              ii++;
            }
        }
      sm.cidx(j+1) = ii;
   }
  // Don't know a-priori the final no of nz.
  sm.maybe_compress ();
\end{lstlisting}
}

Note that both increasing and decreasing the number of non-zero elements in
a sparse matrix is expensive, as it involves memory reallocation. Also as
parts of the matrix, though not its entirety, exist as the old and new copy
at the same time, additional memory is needed. Therefore, if possible this
should be avoided.

\subsection{Using Sparse Matrices in Oct-Files}
\label{sec:octuse}

Most of the same operators and functions on sparse matrices that are
available from the {\Octave} are equally available with {\OctFiles}.
The basic means of extracting a sparse matrix from an \textit{octave\_value}
and returning them as an \textit{octave\_value}, can be seen in the
following example

{\small
\begin{lstlisting}
   octave_value_list retval;

   SparseMatrix sm = 
     args(0).sparse_matrix_value ();
   SparseComplexMatrix scm = 
     args(1).sparse_complex_matrix_value ();
   SparseBoolMatrix sbm = 
     args(2).sparse_bool_matrix_value ();
   ...

   retval(2) = sbm;
   retval(1) = scm;
   retval(0) = sm;
\end{lstlisting}
}

The conversion to an octave-value is automatically handled by the sparse
\textit{octave\_value} constructors, and so no special care is needed.

\section{Conclusion}

This paper has presented the implementation of sparse matrices with
recent versions of Octave. Their storage, creation, fundamental
algorithms used, their implementations and basic operations were also
discussed. Important considerations for the use of sparse matrices
were discussed include efficient manners to create and use them as
well as the return types of several operatoons.

Furthermore, the Octave sparse matrix implementation in {\Octave} version
2.9.5 was compared against {\Matlab} version R14sp2 for the fundamental
addition, multiplication adn left-division operators. It was found that
{\Octave} out-performed {\Matlab} in most cases, with the exceptions
often being for smaller, lower density problems. The efficiency of the
basic {\Octave} sparse matrix implementation has therefore been demonstrated.

Furthermore, we discussed the use of the {\Octave} sparse matrix type
in the context of a real finite element model. The case of a boundary
value Laplace equation, treating the case of a 2D electrically
conductive strip.

Finally, we discussed the use of {\Octave}'s sparse matrices from
within {\Octave}'s dynamically loadable {\OctFiles}. The passing,
means of creating, manipulating and returning sparse matrices
within {\Octave} were discussed. The differences withe {\Octave}'s
\textit{Array$<$T$>$} were discussed.

\bibliographystyle{IEEEbib}
\bibliography{octave2006}

\begin{thebibliography}{10}

\bibitem{octave.website}
John Eaton,
\newblock ``{O}ctave - {A} high-level interactive language for numerical
  computations,'' http://www.octave.org.

\bibitem{eaton.2003}
John Eaton,
\newblock ``Ten years of {O}ctave - {R}ecent developments and plans for the
  future,''
\newblock in {\em Proceedings of the 2nd International Workshop on Distributed
  Statistical Computing}, Vienna, Austria, March 2003, number ISSN 1609-395X.

\bibitem{matlab.gettingstarted}
Mathworks,
\newblock {\em Getting Started with {MATLAB}},
\newblock The Mathworks, Natick NA, 7 edition, 2005.

\bibitem{diffcrash}
Clemens-August Thole and Liquan Mel,
\newblock ``Reasons for scatter in crash simulation results,''
\newblock in {\em NAFEMS Workshop on Use of Stochastics in FEM Analysis},
  Wiesbaden, Germany, May 2003.

\bibitem{saad.sparskit}
Youcef Saad,
\newblock ``{SPARSKIT}: {A} basic tool kit for sparse matrix computation,''
\newblock Tech. {R}ep. version 2, Computer Science Department, University of
  Minnesota, Minneapolis, MN 55455, June 1994.

\bibitem{bondy72}
J.~A. Bondy and U.~S.~R. Murty,
\newblock {\em Graph theory with applications},
\newblock MacMillan, 1972.

\bibitem{gilbert92sparse}
John~R. Gilbert, Cleve Moler, and Robert Schreiber,
\newblock ``Sparse matrices in {MATLAB}: Design and implementation,''
\newblock {\em SIAM Journal on Matrix Analysis and Applications}, vol. 13, no.
  1, pp. 333--356, 1992.

\bibitem{Kahan.1987}
William Kahan,
\newblock {\em Branch Cuts for Complex Elementary Functions, or Much Ado About
  Nothing's Sign Bit}, chapter~7,
\newblock The State of the Art in Numerical Analysis. Clarendon Press, Oxford,
  1987.

\bibitem{eidors.website}
Andy Adler, William Lionheart, and Nick Polydorides,
\newblock ``{EIDORS} - {E}lectrical impedance tomography and diffuse optical
  tomography reconstruction software,'' http://eidors3d.sourceforge.net/.

\bibitem{eidors2006}
Andy Adler and William R~B Lionheart,
\newblock ``Uses and abuses of eidors: An extensible software base for {EIT},''
\newblock {\em Physiological Measurement}, 2006,
\newblock in press.

\bibitem{davis.2006}
Tim Davis,
\newblock ``{CHOLMOD}, a sparse cholesky factorization and modification
  package,''
\newblock {\em ACM Trans. Math. Software}, 2006,
\newblock in preparation.

\bibitem{davis.2004}
Tim Davis,
\newblock ``{UMFPACK} - an unsymmetric-pattern multifrontal method with a
  column pre-ordering strategy,''
\newblock {\em ACM Trans. Math. Software}, vol. 30, no. 2, pp. 196--199, 2004.

\bibitem{ashcraft96applications}
Cleve Ashcraft and Joseph W.~H. Liu,
\newblock ``Applications of {Dulmage-Mendelsohn} decomposition and network flow
  to graph bisection improvement,''
\newblock Tech. {R}ep. CS-96-05, Dept. of Computer Science, York University,
  North York, Ontario, Canada, 1996.

\bibitem{davis.1997}
Tim Davis,
\newblock ``University of florida sparse matrix collection,''
\newblock {\em NA Digest}, vol. 97, no. 23, June 1997,
\newblock http://www.cise.ufl.edu/research/sparse/matrices.

\bibitem{whaley97automatically}
R.~Clint Whaley and Jack~J. Dongarra,
\newblock ``Automatically tuned linear algebra software,''
\newblock Tech. {R}ep. UT-CS-97-366, 1997.

\bibitem{metis.website}
G.~Karypis and V.~Kumar,
\newblock ``"{M}etis: {A} software package for partitioning unstructured
  graphs, partitioning meshes, and computing fill-reducing orderings of sparse
  matrices,'' http://wwwusers.cs.umn.edu/~karypis/metis/metis.html, 1998.

\bibitem{spiel.del_coda}
Cristophe Spiel,
\newblock ``Del {C}oda al {F}ine - pushing {O}ctave's limits,''
  http://octave.sourceforge.net/coda/coda.pdf.

\bibitem{thomas.dal_segno}
Paul Thomas,
\newblock ``Dal {S}egno al {C}oda - the {O}ctave dynamically linked function
  cookbook,'' http://perso.wanadoo.fr/prthomas/intro.html.

\end{thebibliography}

\end{document}